\documentclass[seceq]{ptptex}

\usepackage{graphicx}
\usepackage{epsfig}
\usepackage{wrapft}
\usepackage{here}

\usepackage{bbm}
\usepackage{booktabs}

\newcommand {\Tr}{\mathop{\rm Tr}}

\newcommand {\psibar}{\overline{\psi}}
\newcommand {\Psibar}{\overline{\Psi}}

\newcommand {\dd}{\mbox{d}}

\newcommand {\SO}{\mathop{\rm SO}}
\newcommand {\SU}{\mathop{\rm SU}}

\newcommand{\beq}{\begin{equation}}
\newcommand{\eeq}{\end{equation}}
\newcommand{\beqa}{\begin{eqnarray}}
\newcommand{\eeqa}{\end{eqnarray}}

\title{
Spontaneous breaking of the rotational symmetry\\
in dimensionally reduced super Yang-Mills models
}

\author{
Tatsumi \textsc{Aoyama}$^{1}$,
Jun \textsc{Nishimura}$^{2,3,}$\footnote{E-mail: jnishi@post.kek.jp}
and Toshiyuki \textsc{Okubo}$^{4,}$\footnote{E-mail: tokubo@meijo-u.ac.jp}
}

\inst{
$^1$Kobayashi-Maskawa Institute for the Origin of 
Particles and the Universe (KMI),\\ 
Nagoya University, Nagoya 464-8602, Japan\\
$^2$KEK Theory Center, 
  High Energy Accelerator Research Organization,\\
  1-1 Oho, Tsukuba 305-0801, Japan\\
$^3$The Graduate University for Advanced Studies (SOKENDAI),\\
  1-1 Oho, Tsukuba 305-0801, Japan\\
$^4$Faculty of Science and Technology, Meijo University,
Nagoya 468-8502, Japan
}

\abst{
We investigate the spontaneous breaking of the SO($D$)
symmetry in matrix models, which can be obtained
by the zero-volume limit of pure SU($N$) super Yang-Mills theory 
in $D=6,10$ dimensions.
The $D=10$ case corresponds to 
the IIB matrix model, 
which was proposed as 
a non-perturbative formulation 
of type IIB superstring theory,
and the spontaneous breaking corresponds to 
the dynamical compactification of space-time
suggested in that model.
First we study the $D=6$ case by the Gaussian expansion method,
which turns out to yield clearer results than the previous results
for the $D=10$ case for certain technical reasons.
By comparing the free energy of the SO($d$) symmetric vacua for $d=2,3,4,5$, 
we conclude that the breaking $\SO(6)\to\SO(3)$ actually 
occurs.
We find that the extent of space-time in the shrunken directions
is almost independent of $d$.
In units of this universal scale,
the extended directions seem to have large but still finite extents
depending on $d$.
We show that these results for the extent of space-time
can be explained
quantitatively by an argument based on the low-energy effective theory. 
With these new insights, we reconsider the previous results
for the IIB matrix model,
and find that they are also consistent with
our argument based on the low-energy effective theory.
Thus we arrive at comprehensive
understanding 
and some quantitative predictions concerning
the nature of
the spontaneous symmetry breaking
taking place 
in these models.
The space-time picture that emerges from the IIB matrix model 
and its implication
on possible interpretations
of the model are also discussed.
}

\begin{document}

\maketitle

\section{Introduction}
\label{sec:intro}

It has long been considered that 
matrix models provide a non-perturbative
formulation of string theories analogous to lattice gauge theory
for QCD.
After the revolution
triggered
by the discovery of D-branes,
there appeared concrete proposals for critical 
superstring theories and M 
theory \cite{Banks:1996vh,Ishibashi:1996xs,Dijkgraaf:1997vv},
which take the form of 
large-$N$ reduced models.
For instance, the IIB matrix model \cite{Ishibashi:1996xs},
which is proposed as a non-perturbative formulation of 
type IIB superstring theory, 
can be obtained by the zero-volume limit
of pure $\SU(N)$ super Yang-Mills theory in 10
dimensions.

As a possible non-perturbative phenomenon analogous to quark confinement
in QCD, one can think of the dynamical compactification.
Quarks and gluons, which were introduced by QCD 
as a substructure of hadrons, are considered to be
made invisible due to its own non-perturbative dynamics.
Likewise we might speculate that the 6 extra dimensions, which was
introduced by superstring theory, should be somehow made invisible 
due to its own non-perturbative dynamics.
This speculation has been pursued intensively in the IIB matrix model,
in which the space-time is treated totally as a dynamical object.
The extent of space-time in ten dimensions can be probed
by the eigenvalue distribution of the 10 bosonic 
matrices.\cite{Aoki:1998vn}
If the eigenvalue distribution of the matrices collapses to 
a 4d hypersurface
and the SO(10) symmetry is broken down to SO(4),
it implies that the dynamical compactification to 4d
occurs as a consequence of non-perturbative interactions of superstrings.
The analysis based on the Gaussian expansion 
method (GEM) at the 3rd order \cite{Nishimura:2001sx}
suggested that this might indeed occur.
The free energy of the SO($d$) symmetric vacua is obtained 
for $d=2,4,6,7$, 
and among them $d=4$ is found to give the smallest value.

This result obtained at the 3rd order of GEM
is certainly encouraging\footnote{See also
Refs.~\citen{Kaneko:2005pw,Kaneko:2005kp,%
Itoyama:2009ub,Steinacker:2010rh,Lee:2010zf} 
and references therein
for related works on the appearance of 4d space-time in the IIB matrix model.}, 
and it motivated
5th order \cite{Kawai:2002jk}, 7th order \cite{Kawai:2002ub},
and 8th order \cite{Aoyama:2006rk,Aoyama:2006je} calculations,
which were made possible only after various technical developments.
These works focused on the comparison of the $d=4$ and $d=7$
cases\footnote{See Ref.~\citen{Aoyama:2006di} for a study of the
two cases in a sort of unified framework at the 3rd order.}.
The convergence was seen reasonably in the $d=7$ case, but not quite
in the $d=4$ case.
On the other hand, clear convergence was observed in 
various other matrix models such as
the bosonic IIB matrix model \cite{Nishimura:2002va},
exactly solvable matrix models \cite{Nishimura:2003gz},
and a toy model \cite{Nishimura:2001sq} for the spontaneous breaking 
of the rotational symmetry \cite{Nishimura:2004ts}.
In particular, the last work showed conclusively that
the SO(4) symmetry of the toy model is spontaneously broken
down to SO(2) by calculations up to the 9th order.
This is an explicit example which realizes
the mechanism for the spontaneous symmetry breaking (SSB) 
due to the phase of the complex fermion determinant \cite{NV}.

In this paper we first apply the same method to
the ``six-dimensional version'' of the IIB matrix model,
which can be obtained by the zero-volume limit
of pure $\SU(N)$ super Yang-Mills theory in $D=6$ 
dimensions.\footnote{This model
is termed ``the little IIB matrix model'' in 
Ref.~\citen{Kitazawa:2006hq}
representing the authors' conjecture that it provides
a non-perturbative formulation of 
the (2,0) little string theory \cite{lst} in 6 dimensions.
It would be interesting to consider the implications
of our results in this context.}
This model has a complex fermion determinant,
whose phase has properties similar to those of 
the $D=10$ model
and the toy model mentioned above.
Furthermore, the $D=6$ model is obviously much closer 
to the $D=10$ model
than 
the toy model\footnote{For instance, the toy model
is not supersymmetric. It
has a Gaussian action for the bosonic $N\times N$ matrices, 
and the fermionic variables are introduced
as a fundamental representation of SU($N$) with $N_f$ flavors.
},
and we will see that they seem to share some fundamental properties
concerning the SSB of rotational symmetry. 
On the technical side,
when we apply GEM to the $D=6$ model, the Gaussian action
introduced for the fermionic matrices
involves two types of parameters, 
which transform as a 
self-dual 3-form tensor
and as a vector, respectively.
The former type appears in the study of the $D=10$ model,
whereas the latter type appears in the study of the toy model.
This feature gives us the flexibility in the analysis as seen
in the study of the toy model,
which enables us to obtain results much clearer than
those obtained previously from the study of the $D=10$ model.

For instance, we are able to study
the SO($d$) symmetric vacua 
for $d=5,4,3$ systematically without imposing \emph{ad hoc} symmetries
in the ``extra dimension'' in most cases up to the 5th order of the
expansion.
As a result, we find that the free energy
decreases monotonically as $d$ decreases.
We also obtain the extent of space-time in each of 6 directions.
The extent in the shrunken directions
turns out 
to be 
almost independent of $d$.
In units of this universal scale,
the extent in the extended directions
seem to have large but still finite extents
depending on $d$.
We provide a quantitative explanation of these results
for the extent of space-time
based on 
the low-energy effective theory \cite{Aoki:1998vn,Ambjorn:2000dx}.

We also present some results for the SO(2) symmetric vacuum.
Unfortunately in this case,
we had to impose a discrete symmetry in the extra dimensions
to reduce the number of parameters in the Gaussian action.
However, we do find solutions 
having approximately the universal extent
in the shrunken directions.
Assuming that the universality holds for the SO(2) symmetric vacuum
as well, we find that it has much higher free energy than the other vacua.
This 
is consistent with the 
previous observation
that the two-dimensional space-time is
suppressed by the fermion determinant \cite{NV}.
Thus we conclude that the SO(3) symmetric vacuum 
is chosen dynamically, which implies that 
the SO(6) symmetry breaks down to SO(3) spontaneously
in the $D=6$ model.

Given these new insights,
we reconsider the previous results for the $D=10$ model obtained 
by GEM
up to the 5th order for 
the SO(4) and SO(7) symmetric vacua.
While the results are less clear compared with the $D=6$ case,
the extent in the shrunken directions is similar for the two vacua,
which suggests that the aforementioned universality holds here as well.
On the other hand, the extent in the extended directions
are quite different for the two vacua,
and the values turn out to be consistent 
with the theoretical understanding based on the 
low-energy effective theory we arrive at in this paper.
Assuming that this interpretation of the previous results is correct, 
we predict the extent of space-time
for the SO($d$) symmetric vacua with $d$ other than $d =4,7$.

The rest of this article is organized as follows. 
In Section \ref{sec:model} we define the model and
the observable which serves as an order parameter
of the SSB of SO($D$).
In Section \ref{sec:gaussian} we describe the method we 
use to analyze the model.
In Section \ref{sec:ansatz} we explain the Ansatz
we use to study the SO($d$) symmetric vacua.
In Section \ref{sec:result}
we present the results for 
the SO($d$) symmetric vacua ($d=3,4,5$).
In Section \ref{sec:theo-interpret}
we provide a theoretical understanding of the results
based on the low-energy effective theory.
In Section \ref{sec:so2z4}
we discuss the results for 
the SO(2) symmetric vacuum.
In Section \ref{sec:10dcase} we reconsider
the previous results for the $D=10$ case
from the viewpoint of our understanding based on 
the low energy effective theory.
Section \ref{sec:discussion} is devoted to a summary and
discussions.
In Appendix \ref{sec:symplectic} we describe the notation of
symplectic Majorana-Weyl spinors, which is useful 
for concrete calculations in the $D=6$ model.
In Appendix \ref{sec:KNSresult} we derive the value of the free energy
for the $D=6$ model from a previous analytic work \cite{Krauth:1998xh}
in order to compare
it with the results obtained in the present paper.

\section{The model and the order parameter}
\label{sec:model}

In this paper we study the dimensionally reduced super Yang-Mills
models including the IIB matrix model.
Let us recall that
pure super Yang-Mills theories can be defined
in $D=3$, 4, 6 and 10 dimensions.
By taking the zero-volume limit of each theory,
one obtains matrix models with $D$ bosonic matrices
and their superpartners.
The $D=10$ case corresponds to the IIB matrix model.
The convergence of the partition function was investigated
both numerically \cite{Krauth:1998xh} and analytically \cite{AW}. 
The $D=3$ model is ill-defined since the partition function
is divergent. The $D=4$ model has a real positive
fermion determinant, and Monte Carlo simulation
suggested the absence of SSB \cite{Ambjorn:2000bf}.
(See also Refs.~\citen{Burda:2000mn,Ambjorn:2001xs}.) 
The $D=6$ model and the $D=10$ model both
have a complex fermion determinant,
whose phase is expected to play a crucial 
role \cite{NV,Nishimura:2001sq,Nishimura:2004ts,sign}
in the SSB of SO($D$).

The $D=6$ model can be obtained by the zero-volume limit of 
pure $\SU(N)$ super Yang-Mills theory in six dimensions,
and its partition function is given by 
\begin{eqnarray}
  Z &=& \int \dd A \, \dd \Psi \, \dd \Psibar
\, e^{-S_{\rm b} - S_{\rm f}} \ , 
\label{eq:6dpf} \\
  S_{\rm b} &=& - \frac{1}{4 g^2}  \Tr \, [ A_\mu, A_\nu ]^2  \ , 
\label{eq:sb} \\
  S_{\rm f} &=& - \frac{1}{g^2} 
\Tr\left(\Psibar_\alpha (\Gamma^\mu)_{\alpha\beta}
    [ A_\mu, \Psi_\beta ] \right) \ . 
\label{eq:sf}
\end{eqnarray}
Here $A_\mu$ ($\mu = 1,\cdots,6$) are 
traceless $N\times N$ Hermitian matrices, 
whereas $\Psi_\alpha$ and $\Psibar_\alpha$ ($\alpha = 1,\cdots,4$) are 
traceless $N\times N$ matrices with Grassmannian entries. 
The parameter $g$ can be scaled out by appropriate redefinition
of the matrices, and hence it is just a scale parameter
rather than
a coupling constant.
We therefore
set $g^2 N = 1$ from now on
unless mentioned otherwise.
The integration measure for $A_\mu$, $\Psi_\alpha$ and $\Psibar_\alpha$ 
is given by
\begin{eqnarray}
  \dd A 
  &=& 
  \prod_{a=1}^{N^2-1} \prod_{\mu=1}^{6} \frac{\dd A_\mu^a}{\sqrt{2\pi}} \ , 
\label{eq:measure_a} \\
  \dd \Psi \dd \Psibar 
  &=& 
  \prod_{a=1}^{N^2-1} \, \prod_{\alpha=1}^{4} 
    \dd \Psi_\alpha^a \dd \, \Psibar_\alpha^a \ ,
\label{eq:measure_psi}
\end{eqnarray}
where $A_\mu^a$, $\Psi_\alpha^a$ and $\Psibar_\alpha^a$ are 
the coefficients in the expansion 
$A_\mu = \sum_{a=1}^{N^2-1} A_\mu^a T^a$ etc.\
with respect to the $\SU(N)$ generators $T^a$ 
normalized as $\Tr (T^a T^b) = \frac{1}{2}\delta^{ab}$. 

The model has an $\SO(6)$ symmetry, 
under which $A_\mu$ transforms as a vector, 
and $\Psi_\alpha$, $\Psibar_\alpha$ transform
as Weyl spinors, respectively.
The $4\times 4$ matrices $\Gamma_\mu$ are the gamma matrices after 
the Weyl projection, and their 
explicit form is given, for instance, by 
\begin{align}
  \Gamma_1 &= \sigma_2 \otimes \sigma_2 \ ,
&
  \Gamma_2 &= \sigma_1 \otimes \sigma_2  \ ,
&
  \Gamma_3 &= \sigma_3 \otimes \sigma_2 \ , 
\nonumber \\
  \Gamma_4 &= \mathbbm{1} \otimes \sigma_1 \ ,
&
  \Gamma_5 &= \mathbbm{1} \otimes \sigma_3 \ ,
&
  \Gamma_6 &= -i\, \mathbbm{1} \otimes \mathbbm{1} \ .
\label{eq:gamma}
\end{align}

In order to discuss the spontaneous breaking of the $\SO(6)$ symmetry 
in the large-$N$ limit, 
we consider the ``moment of inertia'' tensor 
\begin{equation}
  T_{\mu\nu} = \frac{1}{N} \Tr (A_\mu A_\nu) \ , 
\label{eq:tmunu}
\end{equation}
which is a $6\times6$ real symmetric tensor. 
Let us represent its eigenvalues as $\lambda_i$ ($i=1,\cdots, 6$) 
with the specific order 
\begin{equation}
  \lambda_1 \geq \lambda_2 \geq \cdots \geq \lambda_6  \ .
\label{eq:lambda}
\end{equation}
If the SO(6) symmetry is not spontaneously broken, 
the expectation values $\langle \lambda_j  \rangle$ 
($j=1, \cdots , 6$) should be all equal in the large-$N$ limit.
Therefore, if we find that they are not equal in the large-$N$ limit,
it implies that the SO(6) symmetry is spontaneously broken.
Thus the expectation values $\langle \lambda_j  \rangle$ 
serve as an order parameter of the SSB.
In Ref.~\citen{NV} 
it was found that the phase of the fermion determinant favors 
such configurations with $d\ge 3$ in which $\lambda_j$ for 
$j=d+1  , \cdots , 6$ 
are much smaller than the others. This suggests the possibility 
that the SO(6) symmetry is broken down to SO($d$) with $d\ge 3$. 
Since the eigenvalue distribution of $A_\mu$ represents 
the extent of space-time in the IIB matrix model \cite{Aoki:1998vn},
the above situation represents
the dynamical compactification to $d$-dimensional space-time.

\section{The Gaussian expansion method}
\label{sec:gaussian}

Since there are no quadratic terms in the actions
(\ref{eq:sb}) and (\ref{eq:sf}),
we cannot perform perturbative expansion in the ordinary sense.
Finding the vacuum of this model is therefore a problem of 
solving a strongly coupled system.
It is known that a certain class of matrix models
can be solved exactly by using various large-$N$ techniques,
but the present model does not belong to 
such a category.
The use of 
GEM
in studying large-$N$ matrix quantum mechanics 
has been advocated by Kabat and Lifschytz \cite{Kabat:2000hp},
and various black hole physics of the dual geometry 
has been discussed \cite{blackholes}.
Applications to simplified versions of the IIB matrix model
were pioneered in Ref.~\citen{Gauss_simpleIIB}.

The starting point of 
GEM
is to introduce a Gaussian term $S_0$ and rewrite the action 
$S=S_{\rm b}+S_{\rm f}$ as 
\begin{equation}
  S = (S_0 +  S) - S_0  \ .
\label{eq:shift_s}
\end{equation}
Then we can perform a perturbative expansion
regarding the first term $(S_0 +  S)$ as
the ``classical action'' and the second term
$(-S_0)$ as the ``one-loop counter term''.
The results at finite order depend, of course, on the choice of the 
Gaussian term $S_0$, which contains many parameters in general. However, 
it is known in various examples that there exists a region of parameters 
in which the results obtained at finite order are almost constant 
with respect to the variation of parameter values. In this ``plateau'' 
region, the dependence on the parameters is considered to 
vanish effectively, 
and the correct result should be reproduced \cite{Stevenson:1981vj}.
Therefore, if one can identify this plateau region, one can make concrete 
predictions. It should be emphasized that the method enables us to obtain 
genuinely \emph{non-perturbative} results, 
although most of the tasks involved 
are equivalent to perturbative calculations. 

There are some cases in which one finds 
more than one plateau regions in the
parameter space.
In that case, each of them 
corresponds to a local
minimum of the effective action, and
the plateau which gives the smallest free energy 
corresponds to the true vacuum. 
These statements have been confirmed explicitly
in exactly solvable matrix models \cite{Nishimura:2003gz}.

As the Gaussian action for the present model,
let us consider the most general one that preserves 
the $\SU(N)$ symmetry.
In order to study the SSB of SO(6),
we have to allow the Gaussian action 
to break the $\SO(6)$ symmetry.
In practice we are going to restrict the parameter space
by imposing a subgroup of $\SO(6)$.
If we find that a plateau region develops for a particular
breaking pattern, we identify it as a local minimum
which breaks the $\SO(6)$ symmetry spontaneously.
By comparing the free energy,
one can determine which local minimum is actually the true vacuum.

Making use of the $\SO(6)$ symmetry of the model,
we can always bring the Gaussian action into the form
\begin{eqnarray}
  S_0 &=& S_{\rm 0b} + S_{\rm 0f} \ , 
\label{eq:gaussian} \\
  S_{\rm 0b} &=& \frac{N}{2} \sum_{\mu = 1}^{6} M_\mu \Tr (A_\mu)^2 \ , 
\label{eq:s0b} \\
  S_{\rm 0f} &=& N \sum_{\alpha,\beta=1}^{4} \mathcal{A}_{\alpha\beta} \Tr 
  ( \Psibar_\alpha \Psi_\beta ) \ ,
\label{eq:s0f}
\end{eqnarray}
where $M_\mu$, $\mathcal{A}_{\alpha\beta}$ are arbitrary parameters. 
The $4\times4$ complex matrix $\mathcal{A}_{\alpha\beta}$ 
can be expanded in terms of the gamma 
matrices as\footnote{In the $D=10$ case,
the vector term in the
fermionic Gaussian action represented by
$m_\mu$ in Eq.~(\ref{eq:decomp_m})
is absent due to the Majorana nature of the fermions.
This difference has an important consequence concerning
the possible Ansatz
to be discussed in Section \ref{sec:ansatz}.
In the $D=10$ case, one cannot have the SO($d$) Ansatz with 
$d=9,8$, while in the $D=6$ case, 
one can have the SO($d$) Ansatz with $d=5,4,3,2$.
The situation for the $D=6$ model is similar to that of the toy model 
\cite{Nishimura:2004ts}, which has the vector term only.
}
\begin{equation}
  \mathcal{A}_{\alpha\beta} 
  = \sum_{\mu = 1}^{6} m_\mu (\Gamma_\mu)_{\alpha\beta}
  + \sum_{\mu,\nu,\rho = 1}^6 \frac{i}{2\cdot 3!} m_{\mu\nu\rho} 
  (\Gamma_{\mu} \Gamma_{\nu}^\dagger \Gamma_{\rho})_{\alpha\beta} \ ,
\label{eq:decomp_m}
\end{equation}
using a vector $m_\mu$ and a self-dual 3-form $m_{\mu\nu\rho}$,
where the self-duality 
\begin{gather}
  m_{\mu\nu\rho} =  \sum_{\kappa,\lambda,\sigma = 1}^6 
  \frac{i}{3!} \, 
\epsilon_{\mu\nu\rho\kappa\lambda\sigma} \,
m_{\kappa\lambda\sigma}
\label{eq:selfdual}
\end{gather}
follows from the Weyl condition for $\Psi_\alpha$. 
Let us then rewrite the partition function (\ref{eq:6dpf}) as
\begin{eqnarray}
  Z &=& Z_0 \,\langle e^{-(S - S_0)} \rangle_0 \ , 
\\
  Z_0 &=& \int \dd A \, \dd \Psi \, \dd \Psibar \, e^{-S_0} \ ,
\end{eqnarray}
where $\langle\,\cdot\,\rangle_0$ is a vacuum expectation value 
with respect to the partition function $Z_0$. 
{}From this one finds that the free energy $F= - \ln Z$ can be expanded 
as 
\begin{eqnarray}
  F &=& \sum_{k = 0}^\infty f_k \ , 
\nonumber \\
  f_0 &=& - \ln Z_0 \ , 
\nonumber \\
  f_k &=& - \sum_{l = 0}^{k} 
  \frac{(-1)^{k-l}}{(k + l)!}\,{}_{k+l}{\rm C}_{k-l}
  \Bigl\langle (S_{\rm b} - S_0)^{k-l} (S_{\rm f})^{2l} 
  \Bigr\rangle_{\rm C,0} 
  \qquad \text{for $k\geq1$} \ ,
\label{eq:f_gem}
\end{eqnarray}
where the subscript ``C'' in $\langle\,\cdot\,\rangle_{\rm C,0}$ 
implies that the connected part is taken. 
The expansion is organized 
so as to
correspond to the loop expansion regarding the insertion
of the 2-point vertex $(-S_0)$ as a contribution from
the one-loop counterterm.
Similarly the expectation value of an observable $\mathcal{O}$ 
can be evaluated as
\begin{eqnarray}
  \langle \mathcal{O} \rangle &=& \langle \mathcal{O} \rangle_0
  + \sum_{k=1}^\infty O_k \ , \nonumber \\
  O_k &=& 
  \sum_{l = 0}^{k} \frac{(-1)^{k-l}}{(k + l)!}\,{}_{k+l}{\rm C}_{k-l}
  \langle 
    \mathcal{O}\,(S_{\rm b} - S_0)^{k-l}\,(S_{\rm f})^{2l} 
  \rangle_{\rm C,0} \  .
\label{eq:o_gem}
\end{eqnarray}

In practice we truncate the series expansion at some finite order.
Then the free energy (\ref{eq:f_gem}) and 
the observable (\ref{eq:o_gem}) 
depend on 
the arbitrary parameters $M_\mu$ and $\mathcal{A}_{\alpha\beta}$
in the Gaussian action.
We search for the values of parameters at which
the free energy becomes stationary by solving 
the ``self-consistency equations''
\begin{equation}
  \frac{\partial}{\partial M_\mu} F = 0 \ , 
\qquad
  \frac{\partial}{\partial m_\mu} F = 0 \ , 
\qquad
  \frac{\partial}{\partial m_{\mu\nu\rho}} F = 0 \ ,
\label{eq:self-consistency}
\end{equation}
and estimate $F$ and $\langle\mathcal{O}\rangle$ 
at the solutions.
As we increase the order of the expansion, the number
of solutions increases. If we find that there are many solutions
close to each other in the parameter space which give similar
results for the free energy and the observables, we may identify
the region as a plateau.

In actual calculation 
it is convenient to derive the series expansion
(\ref{eq:f_gem}) in the following way.
First we consider the action 
\begin{equation}
  \tilde{S} = S_0 + \epsilon \, S_{\rm b} + \sqrt{\epsilon} 
\, S_{\rm f} \ ,
\label{eq:imfa_s}
\end{equation}
and the partition function 
\begin{equation}
  \tilde{Z} = \int \dd A \, \dd \Psi \, 
\dd \Psibar\, e^{-\tilde{S}} \ , 
\label{eq:imfa_pf}
\end{equation}
where $\epsilon$ is a fictitious expansion parameter.
Next we calculate the free energy 
in the $\epsilon$-expansion as
\begin{equation}
  \tilde{F} = - \ln \tilde{Z} = \sum_{k=0}^{\infty} \epsilon^k 
\tilde{f}_{k} \ .
\label{eq:imfa_f}
\end{equation}
Each term $\tilde{f}_{k}$ depends on
the parameters $M_\mu$, $m_\mu$ and $m_{\mu\nu\lambda}$ 
in the Gaussian action $S_0$.
Then we substitute these parameters as
\begin{equation}
  M_\mu \to (1-\epsilon)\,M_\mu \ ,
\qquad
  m_\mu \to (1-\epsilon)\,m_\mu \ ,
\qquad
  m_{\mu\nu\rho} \to (1-\epsilon)\,m_{\mu\nu\rho} \ ,
\label{eq:imfa_improve}
\end{equation}
reorganize the series with respect to $\epsilon$,
and set $\epsilon$ to~1. 
In this way we can reproduce the expression (\ref{eq:f_gem}). 
The action (\ref{eq:imfa_s})
is introduced to obtain the ordinary perturbation theory for 
the first term in (\ref{eq:shift_s}),
and the final step (\ref{eq:imfa_improve})
corresponds to taking account of the second term
in (\ref{eq:shift_s}) as the one-loop counter term.
The main task is to obtain the series (\ref{eq:imfa_f}),
which is exactly what is required for 
ordinary perturbation theory.
We use a similar procedure to obtain the expansion (\ref{eq:o_gem})
for the observables.

Further simplification is possible by 
exploiting the fact that the free energy $F$ is related 
to the two-particle irreducible (2PI) free energy
through the Legendre transformation. 
The number of Feynman diagrams decreases considerably
by the restriction to 2PI diagrams.
This technique, which was introduced in Ref.~\citen{Kawai:2002jk}, 
plays a crucial role in 
performing higher order calculations in this method.

In GEM,
the large-$N$ limit can be taken by simply
drawing Feynman diagrams with the double-line notation,
and keeping only the planar diagrams when evaluating
the free energy (\ref{eq:f_gem}) and 
the observable (\ref{eq:o_gem}).
The whole procedure is automated on a PC
by using a C program.
Calculations at finite $N$ would be more difficult
due to proliferation of diagrams.

In the present model, we use the notation of
symplectic Majorana-Weyl spinors 
(See Appendix~\ref{sec:symplectic} for the details.),
which reduces the number of Feynman diagrams considerably. 
In fact the set of diagrams to be calculated is exactly the same as 
in the IIB matrix model ($D=10$ case),
and hence the list of planar 2PI Feynman diagrams 
in Ref.~\citen{Kawai:2002jk} can be used without modifications.

\section{Ansatz}
\label{sec:ansatz}

There are many arbitrary parameters 
in the Gaussian action (\ref{eq:gaussian}); \emph{i.e.},
one gets
6 from $M_\mu$ and 16 from $\mathcal{A}_{\alpha\beta}$.
(In the case of the $D=10$ model,
one gets 10 from $M_\mu$ and 120 from $\mathcal{A}_{\alpha\beta}$.)
There are as many self-consistency equations as these parameters.
Unfortunately it seems impossible to solve them
in full generality.
However, it is reasonable to expect that some subgroup of SO(6)
symmetry such as SO($d$) with $d =2,3,4,5$ remains unbroken.
This allows us to impose the corresponding symmetry on the 
Gaussian action, and hence to reduce the number of parameters
considerably.
We therefore consider the following Ansatz.
\\ {\bf SO(5) Ansatz (3 parameters)}
\begin{gather}
  M_\mu = (M, M, M, M, M, M_6) \ , 
\qquad
  m_\mu = (0, 0, 0, 0, 0, m_6) \ , 
\qquad
  m_{\mu\nu\rho} = 0 \ .
\nonumber 
\label{eq:so5}
\end{gather}
\\ {\bf SO(4) Ansatz (5 parameters)}
\begin{gather}
  M_\mu = (M, M, M, M, M_5, M_6) \ , 
\qquad
  m_\mu = (0, 0, 0, 0, m_5, m_6) \ , 
\qquad
  m_{\mu\nu\rho} = 0 \ .
\nonumber 
\label{eq:so4}
\end{gather}
\\ {\bf SO(3) Ansatz (8 parameters)}
\begin{gather}
  M_\mu = (M, M, M, M_4, M_5, M_6) \ ,
\nonumber \\
  m_\mu = (0, 0, 0, m_4, m_5, m_6) \ ,
\nonumber \\
  m_{123} = -i m_{456} = \tilde{m} \ , \quad \text{and zero otherwise} .
\nonumber 
\label{eq:so3}
\end{gather}
\\ {\bf SO(2) Ansatz (13 parameters)}
\begin{gather}
  M_\mu = (M, M, M_3, M_4, M_5, M_6) \ ,
\nonumber \\
  m_\mu = (0, 0, m_3, m_4, m_5, m_6) \ ,
\nonumber \\
  m_{123} = -i m_{456} = \tilde{m}_3 \ , 
\quad m_{124} = i m_{356} = - \tilde{m}_4 \ , 
\nonumber \\
  m_{125} = -i m_{346} = \tilde{m}_5 \ , 
\quad m_{126} = i m_{345} = - \tilde{m}_6 \ ,
\quad \text{and zero otherwise} .
\nonumber 
\label{eq:so2}
\end{gather}

In some cases it turns out that the obtained solution
has extra symmetries.
For instance, some solutions obtained with the $\SO(4)$ Ansatz
satisfy $M_5=M_6$ and $m_5=m_6$.
This implies that in fact the solutions have a larger symmetry
$\SO(4) \times \mathbb{Z}_2$,
where $\mathbb{Z}_2$ corresponds to 
exchanging 5th and 6th directions.
(By this we actually mean
$x_5 \mapsto x_6$ and $x_6 \mapsto x_5$ combined with
$x_4 \mapsto - x_4$ in order to make the transformation 
an element of the SO(6) group. Note that this is \emph{not} 
equivalent to imposing symmetry under a 90-degree rotation 
$x_5 \mapsto x_6$ and $x_6 \mapsto -x_5$.)

Similarly, solutions obtained with the $\SO(3)$ Ansatz
can have larger symmetries. Here we list only the symmetries
that we have encountered in actual calculations.
(i) $\SO(3) \times \mathbb{Z}_2$,
where the meaning of $\mathbb{Z}_2$ is the same as in
the $\SO(4) \times \mathbb{Z}_2$ case.
The solution then satisfies 
$M_5= M_6$, $m_5=m_6$ and $m_4=0$, 
which leaves us with 5 parameters.
(ii) $\mbox{SO(3)} \times \mathbb{Z}_3$,
where $\mathbb{Z}_3$ corresponds to 
cyclically permuting the 4th, 5th and 6th directions.
The solution then satisfies
$M_4= M_5 = M_6$ and $m_4=m_5=m_6$, which leaves us
with 4 parameters.
(iii) $\SO(3) \times \SO(2)$,
where the $\SO(2)$ corresponds to a rotation
involving the 4th and 5th directions, for instance.
The solution then satisfies
$M_4= M_5$ and $m_4=m_5=0$,
which leaves us with 5 parameters.
(iv) $\SO(3) \times \SO(3)$.
The solution satisfies
$M_4= M_5 = M_6$ and $m_4=m_5=m_6=0$,
which leaves us with 3 parameters.

With the available computational resources, 
the order 5 calculation was possible
with no more than 5 parameters.
We therefore impose
either the $\mbox{SO(3)} \times \mathbb{Z}_3$ symmetry
or the $\SO(3) \times \SO(2)$ symmetry to study the
SO(3) symmetric vacuum at the 5-th order.\footnote{The 
solutions obtained in this way can also have
$\SO(3) \times \SO(3)$ symmetry.
We classify such solutions separately.}
Except for this particular case, we are able to 
study the SO($d$) symmetric vacua for $d=3,4,5$
without imposing \emph{ad hoc}
symmetries in the extra dimensions.
This turns out to be a big advantage compared with the previous studies
for the $D=10$ model \cite{Nishimura:2001sx,%
Kawai:2002jk,Kawai:2002ub,%
Aoyama:2006rk,Aoyama:2006je,Aoyama:2006di}.

Unfortunately, we could not study 
the SO(2) symmetric vacuum without imposing extra symmetries.
We therefore decided to
impose the $\mbox{SO(2)} \times \mathbb{Z}_4$ symmetry, where
$\mathbb{Z}_4$ corresponds to cyclically permuting 
the 3rd, 4th, 5th and 6th directions.
(By this we actually mean
$x_3 \mapsto x_4$, $x_4 \mapsto x_5$,
$x_5 \mapsto x_6$ and $x_6 \mapsto x_3$ combined with
$x_1 \mapsto - x_1$ in order to make the transformation 
an element of the SO(6) group.)
In that case the solution satisfies $M_3 = M_4 = M_5 = M_6$,
$m_3 = m_4 = m_5 = m_6$ and 
$\tilde{m}_3 =  \tilde{m}_4 = \tilde{m}_5 =  \tilde{m}_6$,
which leaves us with 4 parameters.
The results turn out to be
somewhat more subtle than those
for the SO($d$) symmetric vacua ($d=3,4,5$),
possibly due to the above restriction.
We therefore discuss the SO(2) symmetric vacuum separately
in Section \ref{sec:so2z4}.

\section{Results for the SO($d$) symmetric vacua ($d=3,4,5$)}
\label{sec:result}

\begin{figure}[b]
\centerline{\includegraphics[width=.8\textwidth]{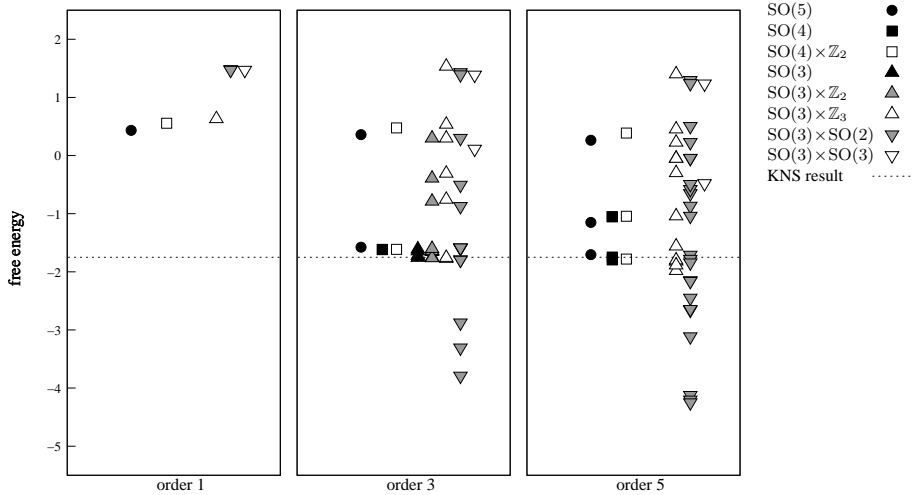}}
\caption{%
The free energy density (\ref{eq:fedensity})
evaluated at solutions of the self-consistency equations 
at orders 1,3 and 5.
Each symbol represents the largest symmetry that the solution has.
The dotted line represents the value ($-\frac{7}{4}$) 
obtained from the KNS conjecture. }
\label{fig:f}
\end{figure}

For each Ansatz,
we first obtain the free energy 
up to the 5th order
as a function of the parameters in the Gaussian action.
By differentiating the free energy with respect to the
parameters, we obtain
the self-consistency equations,
which we solve numerically by Mathematica.
The free energy evaluated at each solution is
plotted in Fig.\ \ref{fig:f}
for the SO($d$) Ansatz ($d=3,4,5$) described 
in the previous Section.
More precisely, we plot 
``the free energy density'' defined as
\begin{equation}
  f = 
  \lim_{N\to\infty} \left\{
  \frac{F}{N^2 - 1} 
  - \left( -\ln N + \ln 2 + 1 \right) 
  \right\} \ ,
\label{eq:fedensity}
\end{equation}
in terms of the free energy $F = - \ln Z$ of the $D=6$ model
(\ref{eq:6dpf}).
As we explained in Section \ref{sec:ansatz},
solutions can have larger symmetries than the one imposed
by the Ansatz.
We classify the solutions for each Ansatz
by the largest symmetry they actually have.

\begin{table}[b]
\begin{center}
\begin{tabular}{c l  lllll c}
\toprule
order & 
symmetry & 
\makebox[5em]{$f$} &
\makebox[3.5em]{$\langle \lambda_{1,2,3} \rangle$} &
\makebox[3.5em]{$\langle \lambda_4 \rangle$} &
\makebox[3.5em]{$\langle \lambda_5 \rangle$} &
\makebox[3.5em]{$\langle \lambda_6 \rangle$} &
\\
\midrule
1 & $\SO(5)$                       & $ \phantom{-}0.43323$ & $ 0.46291$ & 
\multicolumn{1}{c}{--} & \multicolumn{1}{c}{--} & $ 0.15430$ & $\ast$ \\
\cline{2-8}
  & $\SO(4)\!\times\!\mathbb{Z}_2$ & $ \phantom{-}0.55517$ & $ 0.50813$ & 
\multicolumn{1}{c}{--} & $ 0.31861$ & $ 0.12498$ & $\ast$ \\
\cline{2-8}
  & $\SO(3)\!\times\!\mathbb{Z}_3$ & $ \phantom{-}0.63196$ & $ 0.55965$ & 
$ 0.36511$ & 
\multicolumn{1}{c}{--}
& $ 0.07153$ & $\ast$ \\
\cline{2-8}
  & $\SO(3)\!\times\!\SO(2)$       & $ \phantom{-}1.48083$ & $ 0.14434$ & 
$ 0.86603$ & \multicolumn{1}{c}{--} & $ 0.43301$ & \\
\cline{3-8}
  &                                & $ \phantom{-}1.46896$ & $ 0.13949$ & 
$ 0.73581$ & \multicolumn{1}{c}{--} & $ 0.67952$ & \\
\cline{2-8}
  & $\SO(3)\!\times\!\SO(3)$       & $ \phantom{-}1.46897$ & $ 0.71685$ & 
$ 0.13950$ & \multicolumn{1}{c}{--} & \multicolumn{1}{c}{--} & \\
\hline
3 & $\SO(5)$                       & $-1.57681$ & $ 0.69925$ & 
\multicolumn{1}{c}{--} & \multicolumn{1}{c}{--} & $ 0.18720$ & $\ast$ \\
\cline{2-8}
  & $\SO(4)$                       & $-1.61405$ & $ 0.87262$ & 
\multicolumn{1}{c}{--} & $ 0.26392$ & $ 0.18597$ & $\ast$ \\
\cline{3-8}
  &                                & $-1.61942$ & $ 0.51325$ & 
\multicolumn{1}{c}{--} & $ 2.07517$ & $ 0.18295$ & \\
\cline{2-8}
  & $\SO(4)\!\times\!\mathbb{Z}_2$ & $-1.61396$ & $ 0.87559$ & 
\multicolumn{1}{c}{--} & $ 0.25981$ & $ 0.18587$ & $\ast$ \\
\cline{2-8}
  & $\SO(3)$                       & $-1.60777$ & $ 0.33740$ & 
$ 6.98126$ & $ 0.11361$ & $ 0.11036$ & \\
\cline{3-8}
  &                                & $-1.64285$ & $ 0.59561$ & 
$ 2.58941$ & $ 0.20836$ & $ 0.17600$ & \\
\cline{3-8}
  &                                & $-1.72044$ & $ 0.65091$ & 
$ 1.84641$ & $ 0.33620$ & $ 0.18244$ & \\
\cline{3-8}
  &                                & $-1.75426$ & $ 0.67863$ & 
$ 2.43441$ & $ 0.18529$ & $ 0.17919$ & \\
\cline{2-8}
  & $\SO(3)\!\times\!\mathbb{Z}_2$ & $-1.60108$ & $ 0.35849$ & 
$ 6.37253$ & $ 6.37253$ & $ 0.12631$ & \\
\cline{3-8}
  &                                & $-1.63294$ & $ 1.02317$ & 
$ 0.55197$ & $ 0.24407$ & $ 0.18307$ & \\
\cline{3-8}
  &                                & $-1.64022$ & $ 0.61060$ & 
$ 2.31060$ & $ 0.23485$ & $ 0.07816$ & \\
\cline{3-8}
  &                                & $-1.65519$ & $ 0.28107$ & 
$ 9.59445$ & $ 0.08448$ & $ 0.07992$ & \\
\cline{3-8}
  &                                & $-1.74649$ & $ 0.68185$ & 
$ 2.17747$ & $ 0.21743$ & $ 0.18003$ & \\
\cline{3-8}
  &                                & $-1.76129$ & $ 1.39517$ & 
$ 0.21393$ & $ 0.19198$ & $ 0.17947$ & $\ast$ \\
\cline{2-8}
  & $\SO(3)\!\times\!\mathbb{Z}_3$ & $-1.75564$ & $ 1.37852$ & 
$ 0.21601$ & 
\multicolumn{1}{c}{--}
& $ 0.16208$ & $\ast$ \\
\cline{3-8}
  &                                & $-1.77159$ & $ 1.43148$ & 
$ 0.20030$ & 
\multicolumn{1}{c}{--}
& $ 0.15052$ & $\ast$ \\
\cline{2-8}
  & $\SO(3)\!\times\!\SO(2)$       & $-1.58507$ & $ 0.42282$ & 
$ 1.29610$ & \multicolumn{1}{c}{--} & $ 0.17334$ & \\
\cline{3-8}
  &                                & $-1.58596$ & $ 0.49087$ & 
$ 1.10749$ & \multicolumn{1}{c}{--} & $ 0.18127$ & \\
\cline{3-8}
  &                                & $-1.59755$ & $ 0.88526$ & 
$ 0.47466$ & \multicolumn{1}{c}{--} & $ 0.18349$ & \\
\cline{3-8}
  &                                & $-1.79126$ & $ 1.47103$ & 
$ 0.18007$ & \multicolumn{1}{c}{--} & $ 0.18387$ & $\ast$ \\
\cline{3-8}
  &                                & $-1.79828$ & $ 1.51783$ & 
$ 0.16939$ & \multicolumn{1}{c}{--} & $ 0.18120$ & $\ast$ \\
\hline
5 & $\SO(5)$                       & $-1.70472$ & $ 0.78386$ & 
\multicolumn{1}{c}{--} & \multicolumn{1}{c}{--} & $ 0.20789$ & $\ast$ \\
\cline{2-8}
  & $\SO(4)$                       & $-1.74598$ & $ 0.55629$ & 
\multicolumn{1}{c}{--} & $ 2.78951$ & $ 0.19406$ & \\
\cline{3-8}
  &                                & $-1.79599$ & $ 1.11197$ & 
\multicolumn{1}{c}{--} & $ 0.20256$ & $ 0.18450$ & $\ast$ \\
\cline{2-8}
  & $\SO(4)\!\times\!\mathbb{Z}_2$ & $-1.78072$ & $ 1.01665$ & 
\multicolumn{1}{c}{--} & $ 0.27056$ & $ 0.20861$ & $\ast$ \\
\cline{2-8}
  & $\SO(3)\!\times\!\mathbb{Z}_3$ & $-1.55969$ & $ 1.83885$ & 
$ 0.19760$ & 
\multicolumn{1}{c}{--}
& $ 0.14390$ & \\
\cline{3-8}
  &                                & $-1.79936$ & $ 1.78423$ & 
$ 0.16467$ & 
\multicolumn{1}{c}{--}
& $ 0.17362$ & $\ast$ \\
\cline{3-8}
  &                                & $-1.81743$ & $ 1.67816$ & 
$ 0.16476$ & 
\multicolumn{1}{c}{--}
& $ 0.14599$ & $\ast$ \\
\cline{2-8}
  & $\SO(3)\!\times\!\SO(2)$       & $-1.71169$ & $ 0.71671$ & 
$ 0.90525$ & \multicolumn{1}{c}{--} & $ 0.20783$ & \\
\cline{3-8}
  &                                & $-1.78197$ & $ 1.77827$ & 
$ 0.32212$ & \multicolumn{1}{c}{--} & $ 0.11537$ & \\
\cline{3-8}
  &                                & $-1.84330$ & $ 1.70995$ & 
$ 0.14987$ & \multicolumn{1}{c}{--} & $ 0.17730$ & $\ast$ \\
\bottomrule
\end{tabular}
\end{center}
\caption{%
Numerical values of the free energy density and the eigenvalues 
of the moment-of-inertia tensor evaluated at the solutions 
that appear in Fig.\ \ref{fig:f} for order 1,
and at the solutions that appear in Fig.\ \ref{fig:f2} for orders 3 and 5.
The dash in some columns for $\langle \lambda_i \rangle$
indicates that the value is the same as the one on the left column
due to the symmetry of the solution.
Asterisks in the right-most
column indicate the solutions we consider to be ``physical''.
}
\label{tbl:result}
\end{table}

\begin{figure}[b]
\centerline{\includegraphics[width=.8\textwidth]{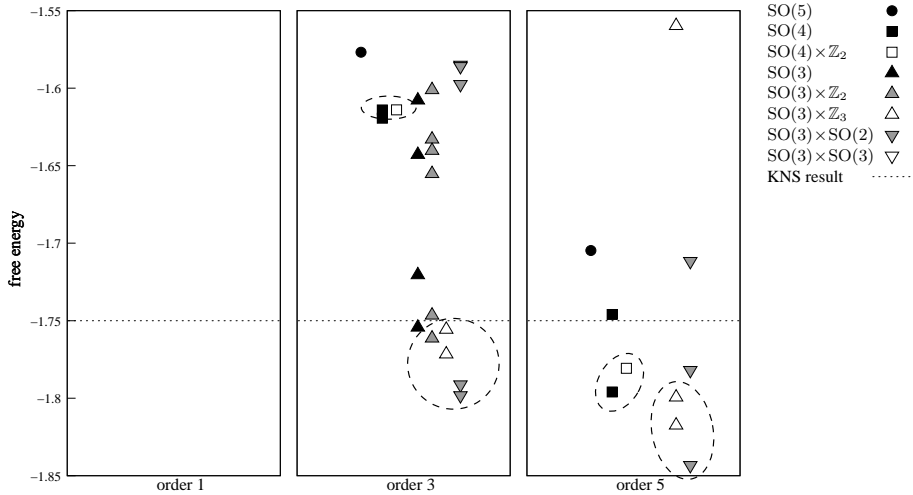}}
\caption{%
The same as Fig.\ \ref{fig:f} except that we zoom up
a small region of the free energy density near the KNS value.
The data points surrounded by the dashed lines correspond
to the concentrating solutions, which indicate the formation of plateaus.
}
\label{fig:f2}
\end{figure}

The horizontal dotted line represents the result
$f=-\frac{7}{4}=-1.75$ obtained in Appendix \ref{sec:KNSresult} 
from the 
analytic formula for the partition function conjectured 
by Krauth, Nicolai and Staudacher \cite{Krauth:1998xh} (KNS) 
combining their Monte Carlo results at small $N$
and earlier analytic works \cite{Green:1997tn,Moore:1998et}.
As we go to higher orders, we obtain many solutions
with the free energy density close to the KNS value.
We take this as an evidence for the validity of
GEM
in the present model.
In Fig.\ \ref{fig:f2} we zoom up the region near the 
KNS value. Table~\ref{tbl:result} shows
the numerical values obtained for the free energy density
at each solution.
For orders 3 and 5,
we restrict the solutions to those displayed
in Fig.\ \ref{fig:f2}.

In Table~\ref{tbl:result} we also
present the six eigenvalues 
of the moment-of-inertia tensor (\ref{eq:tmunu}),
which enable us to probe the extent of space-time 
in each direction.
To obtain these values,
we expand the expectation value of the observable (\ref{eq:tmunu})
by using the formula (\ref{eq:o_gem}) and evaluate it
at the solution of the self-consistency 
equations (\ref{eq:self-consistency}).
Note that the result obtained in this way is not
necessarily diagonal.
For the $\SO(4)$ Ansatz, for instance, 
it takes the form
\begin{equation}
 \langle  T_{\mu\nu} \rangle 
= \left\langle \frac{1}{N}\Tr (A_\mu A_\nu) \right\rangle =
  \begin{pmatrix}
    C & &        &   & \vline \\
      & C &  &   & \vline \\
      &   & C &   & \vline \\
      & &        & C & \vline \\
    \hline 
      & &        &   & \vline & c_1 & c_3 \\
      & &        &   & \vline & c_3 & c_2 \\
  \end{pmatrix} \ .
\label{eq:tmunu:so4}
\end{equation}
In such a case, we have to diagonalize the matrix to 
obtain the eigenvalues.

There are some solutions in Table~\ref{tbl:result} 
having smaller values
in the directions involved in the preserved $\SO(d)$ symmetry
than in the remaining directions.
Such solutions are discarded and will not be considered
in what follows.
At orders 3 and 5, we find a set of solutions for $d=3,4$
giving similar values for 
the free energy density and the eigenvalues 
$\langle \lambda_i \rangle$.
We consider this as the concentration of solutions \cite{Kawai:2002jk},
which indicates the formation of plateaus in the space of parameters
explained in Section \ref{sec:gaussian}.
The data points in Fig.\ \ref{fig:f2} corresponding 
to such solutions are surrounded by the dashed lines.
Thus we can pick up the ``physical solutions'' for each $d$
without much ambiguity,
which we have marked by asterisk in the right-most column 
of Table~\ref{tbl:result}.

\begin{figure}[t]
\centerline{\includegraphics[width=.8\textwidth]{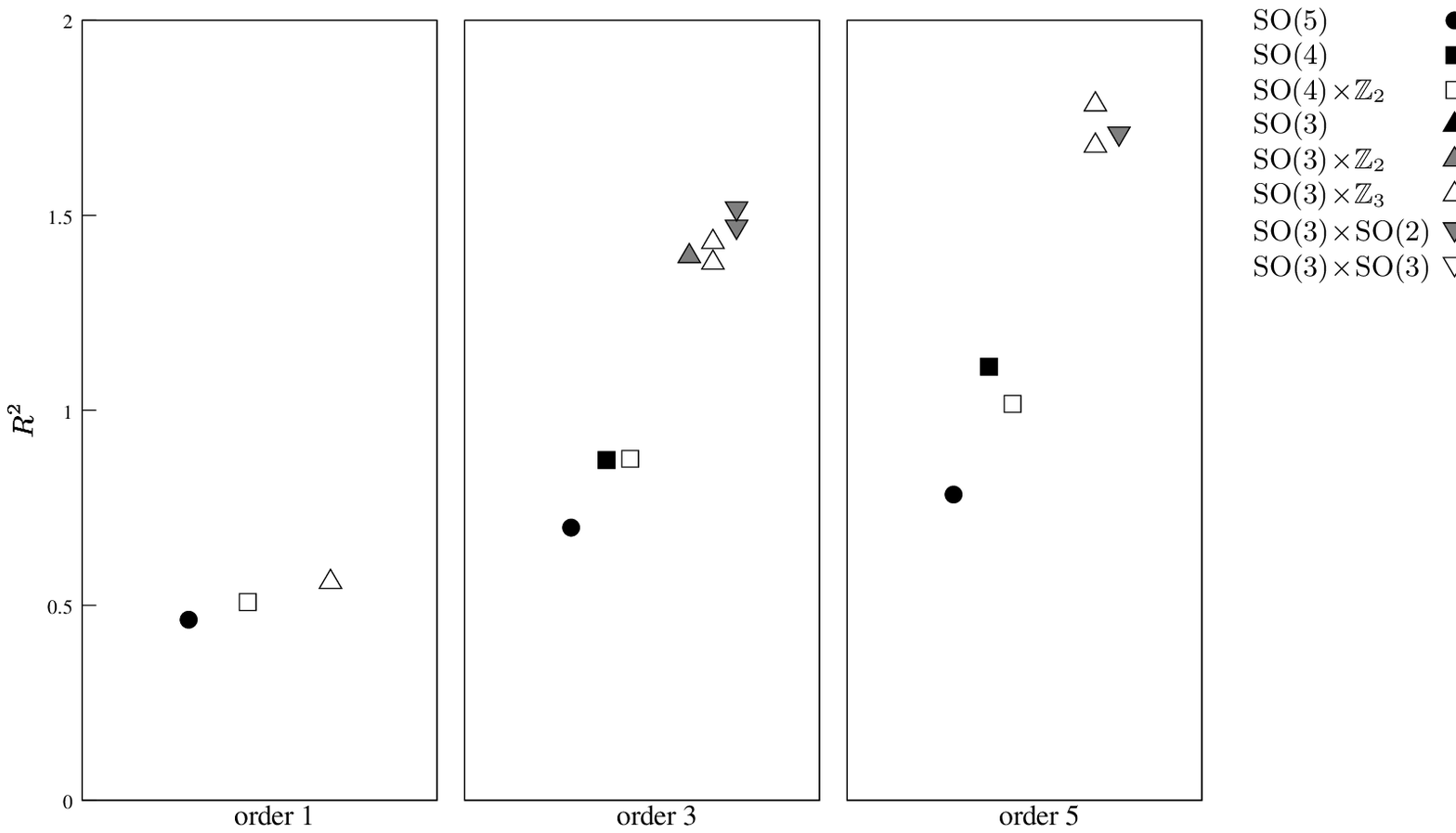}}
\centerline{\includegraphics[width=.8\textwidth]{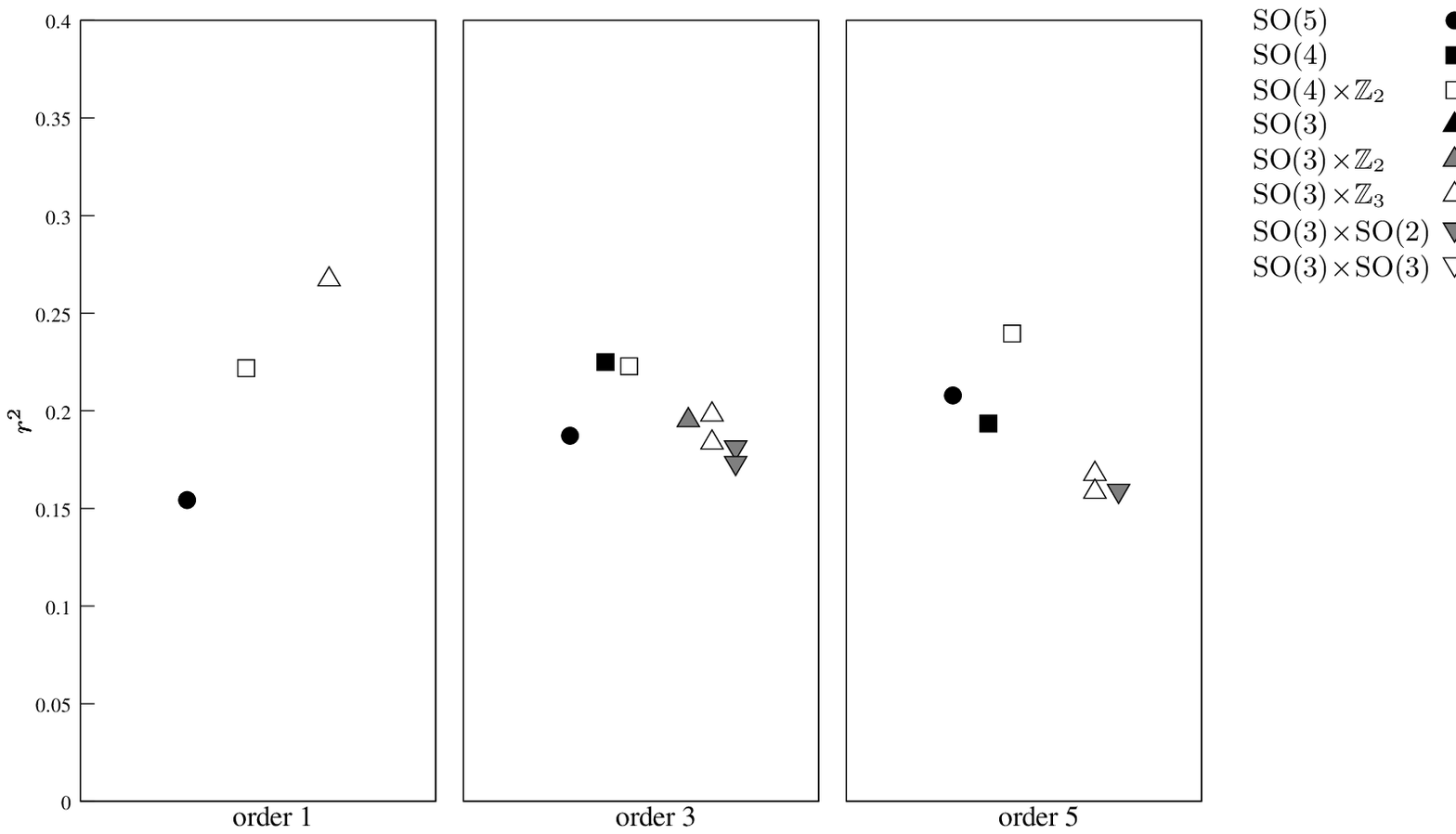}}
\caption{%
The extent of space-time $R^2$ and $r^2$ in the extended (upper panel)
and shrunken directions (lower panel), respectively, evaluated
at the ``physical solutions'' that are marked by
asterisk in Table~\ref{tbl:result}. 
}
\label{fig:r}
\end{figure}

The free energy density 
obtained by averaging over the physical solutions for each $d=3,4,5$
at orders 3 and 5
is found to decrease monotonically as $d$ decreases.
The deviation from the KNS value is reasonable
considering 
the accuracy of the present method at this order.

It is interesting to note in Table~\ref{tbl:result}
that all the SO(3) symmetric solutions without extra symmetries
found at order 3 are actually the ones we decided to discard
because of having larger extent in the extra dimensions.
Therefore, the fact that
we could not study the SO(3) symmetric vacuum at order 5
without imposing extra symmetries actually may not be that harmful
in the present analysis.
On the other hand, we find in
Fig.\ \ref{fig:f}
that the $\SO(3)\times \SO(3)$ symmetric solutions 
have larger free energy density
and hence do not appear in Fig.\ \ref{fig:f2}.
This is understandable
since the $\SO(3)\times \SO(3)$ symmetry
forces all the $m_\mu$'s to vanish,
and therefore the situation can be
qualitatively different from 
the other SO(3) symmetric solutions,
which have some non-zero components in $m_\mu$.

\begin{figure}[t]
\centerline{\includegraphics[width=.8\textwidth]{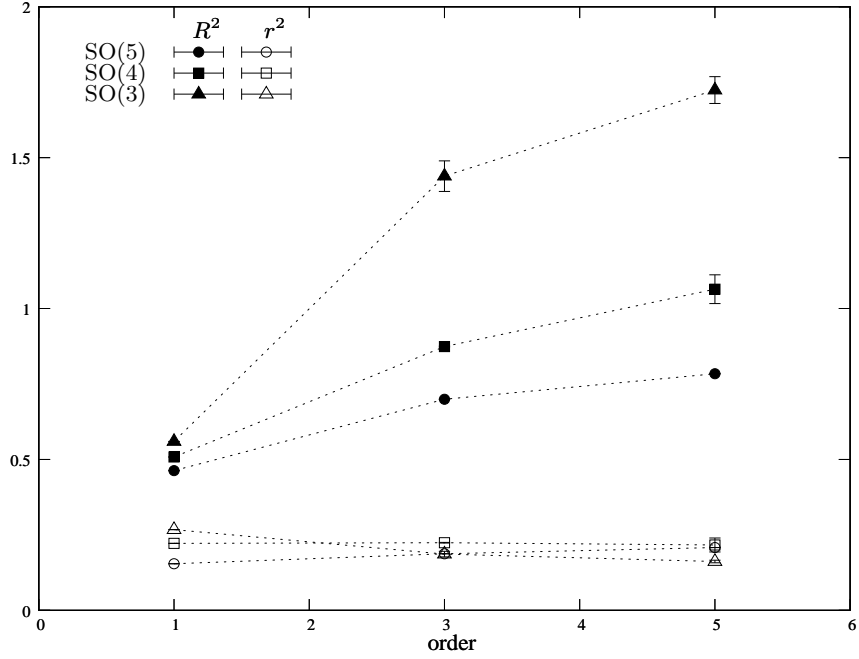}}
\caption{%
The extent of space-time $R^2$ and $r^2$ 
in the extended
and shrunken directions, respectively, 
are plotted at each order by taking an average
over the ``physical solutions'' for each Ansatz.
}
\label{fig:rr}
\end{figure}

Let us also discuss the results for the extent of space-time.
For the $\SO(d)$ Ansatz, 
the $d$ large eigenvalues
of $ \langle  T_{\mu\nu} \rangle $
are equal due to the imposed SO($d$) symmetry,
and we denote the value as $R^2$.
The remaining $(6-d)$ eigenvalues for each ``physical solution''
turn out to be quite close to each other 
and we denote the mean value as $r^2$.
In Fig.\ \ref{fig:r} we 
plot the values of $R^2$ and $r^2$ evaluated 
at the ``physical solutions'' for each Ansatz. 
(Note that the scale is different
for the upper and lower panels.)
Taking the average within each Ansatz, we obtain 
the plot in Fig.\ \ref{fig:rr}.
We put error bars representing the mean square error 
when there are more than one physical solutions 
within the Ansatz.

We find that $r^2$ is quite stable
against increasing the order of the expansion,
and it seems to have an approximately universal value $r^2 \sim 0.2$
for all the SO($d$) symmetric vacua with $d=3,4,5$.
On the other hand, we find that $R^2$ for each $d$
increases with the order, and the convergence is not clear
from this plot alone.
However, we observe a clear tendency at each order that 
$R^2$ is larger for smaller $d$ in contrast to
the universal behavior of $r^2$.

\section{Interpretation based on the low-energy effective theory}
\label{sec:theo-interpret}

In this Section we provide a theoretical 
understanding of the results in the previous Section
based on the low-energy effective theory.
Let us start from the action (\ref{eq:sb}) and (\ref{eq:sf}),
where we keep the scale parameter $g$ 
unspecified 
throughout this Section 
(instead of fixing it by $g^2 N =1$ as we have been doing so far) 
to make some arguments clearer.
First we decompose the bosonic and fermionic matrices as
\beqa
\label{decompose-diag-off}
(A_\mu)_{ij} &=& x_{i \mu} \delta_{ij} + a_{\mu ij}  \ , \\
(\Psi_\alpha)_{ij} &=& \xi_{i \alpha} \delta_{ij} 
+ \varphi_{\alpha ij}  \ , \\
(\bar{\Psi}_\alpha)_{ij} &=& \bar{\xi}_{i \alpha} \delta_{ij} 
+ \bar{\varphi}_{\alpha ij}  \ ,
\label{decompose-diag-off2}
\eeqa
where $a_{\mu ij}$, $\varphi_{\alpha ij}$ and 
$\bar{\varphi}_{\alpha ij}$
contain only off-diagonal elements.
We may view $x_{i \mu}$ as the $\mu$-th coordinate of a point 
$\vec{x}_i$
in $6$-dimensional space-time.
When $\sqrt{( \vec{x}_i - \vec{x}_j )^2} \gg \sqrt{g}$, namely when
all the $N$ points are separated from each other, 
we can integrate out the off-diagonal parts
$a_{\mu ij}$, $\varphi_{\alpha ij}$ and $\bar{\varphi}_{\alpha ij}$ 
at one loop
to obtain the effective action for the diagonal elements
$x_{i \mu}$, $\xi_{i \alpha}$ and $\bar{\xi}_{i \alpha}$.
If one sets $\xi_{i \alpha}=\bar{\xi}_{i \alpha}=0$, 
the effective action actually
vanishes as a consequence of supersymmetry.
However, the integration over $\xi_{i \alpha}$ and $\bar{\xi}_{i \alpha}$
induces a branched-polymer-like interaction 
for $x_{i \mu}$ \cite{Aoki:1998vn}.

Let us briefly review this calculation.
Adding the gauge fixing term 
corresponding to the Feynman gauge, one obtains
the bosonic action relevant at one loop as
\beq
S_{\rm b,1-loop} =
\frac{1}{g^2}
\sum_{i<j}  ( \vec{x}_{ i } - \vec{x}_{ j}  )^2  | a_{\mu ij} | ^2  \  .
\label{bosonic-OL}
\eeq
If we set $\xi_{i \alpha}=0$, 
the fermionic action relevant at one loop becomes
\beq
S_{\rm f,1-loop}(\xi=\bar{\xi}=0) =
- \frac{1}{g^2}
(\Gamma_\mu)_{\alpha\beta}
\sum_{i \neq j}  ( x_{ i \mu } - x_{ j \mu }  )  
\bar{\varphi}_{\alpha ji} \varphi_{\beta ij}  \  .
\label{fermionic-OL}
\eeq
Let us first integrate over
the bosonic variables ($a_{\mu ij}$).
Including the factor $\{\Delta (x) \}^{2}$ coming
the Faddeev-Popov determinant associated with the gauge fixing, where
\beq
\Delta (x) = \prod _{i<j} ( \vec{x}_{ i } - \vec{x}_{ j}  )^2  \ ,
\eeq
we obtain a factor $\{\Delta (x) \}^{-4}$,
which represents an attractive potential between every pair of $\vec{x}_i$.
We can easily see that the integration over
the fermionic 
variables ($\varphi_{\alpha ij}$ and $\bar{\varphi}_{\alpha ij}$)
yields a factor $\{\Delta (x) \}^{4}$, which exactly
cancels the attractive potential induced by the bosonic variables.

Let us then consider what happens if we do not set 
$\xi_{i \alpha}=\bar{\xi}_{i \alpha}=0$.
Here we follow the formulation in Ref.~\citen{Ambjorn:2000dx}.
The fermionic part of the action (\ref{eq:sf})
relevant at one loop reads
\beqa
S_{\rm f, 1-loop} &=&
- \frac{1}{g^2} \, 
  (\Gamma_\mu)_{\alpha\beta}
\sum_{i \neq j} \Bigl\{
 ( x_{ i \mu} - x_{ j \mu}  )  \bar{\varphi}_{\alpha ji}
\varphi_{\beta ij}  \nonumber \\
&~&   \quad \quad \quad \quad \quad \quad
- \bar{\varphi}_{\alpha ji} 
( \xi_{ i \beta} - \xi_{ j \beta}  ) a_{\mu ij}  
-  a_{\mu ji}  ( \bar{\xi}_{ i \alpha} - \bar{\xi}_{ j \alpha  }  )
\varphi_{\beta ij}  \Bigl\}  \  .
\label{Sf-decomp}
\eeqa
Completing the square with respect to $\varphi _{\alpha ij}$
and $\bar{\varphi} _{\alpha ij}$,
and integrating over them, one obtains a factor
$\{\Delta (x) \}^{4}$ we encountered above.
The remaining term in the action is given by
\beq
\tilde{S}_{\rm f, 1-loop} =
-   \frac{1}{g^2} \, \sum _{i  j }
\bar{\xi}_{i \alpha}   M _{i\alpha , j\beta} \xi _{j\beta}  \ , 
\label{action_quad_ferm}
\eeq
where
$M _{i\alpha , j\beta}$ is a $4N$ $\times$ $4N$ matrix
given as
\beqa
M _{i \alpha , j\beta}
&=&  \frac{(x_{i\rho} - x_{j\rho})}{(\vec{x}_i - \vec{x}_j)^2}
( \Gamma _\mu \Gamma _{\rho}^\dag 
\Gamma _\sigma )_{\alpha\beta}
( a_{\mu ji}  a_{\sigma ij} - a_{\mu ij}  a_{\sigma ji} )
~~~~~ \mbox{for}~~~i \neq j  \ , 
\label{detM}
\\
M _{i \alpha , i\beta} &=&
- \sum _{j\neq i} M _{i \alpha , j\beta}  \  .
\label{defMprime}
\eeqa
Integration over $\xi_{i\alpha}$ and $\bar{\xi}_{i\alpha}$ gives
the determinant\footnote{Strictly speaking, we have to 
project out the zero mode $\sum_i \xi_{i \alpha}$ 
and $\sum_i \bar{\xi}_{i \alpha}$,
which corresponds to the trace part of
the fermionic matrices $\Psi_\alpha$ and $\bar{\Psi}_\alpha$
as described in Ref.~\citen{Ambjorn:2000dx}. 
We omit this detail since it is not relevant to 
the arguments below.}
$\det \, M $, which
represents an attractive force between 
$x_{i\mu}$ and $x_{j \nu}$ connected by a bond 
in a branched polymer as one can see from 
the definition of determinants.
In order for the one-loop approximation to be valid,
we have to impose
$|\vec{x}_i - \vec{x}_j| > a_{\rm cut}$ for all $i \neq j$.
The cutoff $a_{\rm cut} \sim \sqrt{g}$ 
is expected to appear dynamically through
the non-perturbative effects of the 
off-diagonal elements \cite{Ambjorn:2000bf}.
The bond length 
of the branched polymer is
given approximately by $a_{\rm bond}\sim a_{\rm cut}$.

The determinant
$\det \, M $ is complex,
and the phase factor actually plays an important role 
in the SSB of SO(6).
However, let us for the moment consider the dynamics
of a system with $|\det \, M|$ omitting the phase.
Due to general properties of a branched polymer,
the extent of the distribution of points 
is given by $\ell=O(a_{\rm bond} N^{1/4})$, 
where 
$N$ gives the number of points in the branched polymer.
Thus we obtain
$\ell  \sim \sqrt{g} N^{1/4}$ as first predicted in 
Ref.~\citen{Aoki:1998vn}.
This property has been confirmed by Monte Carlo simulations
for the $D=6,10$ models \cite{Ambjorn:2000dx} 
using the one-loop approximation,
and for the $D=4$ model \cite{Ambjorn:2000bf}
\emph{without} using the one-loop approximation.

As we mentioned in Section \ref{sec:model},
the phase of the determinant favors collapsed configurations,
and the branched-polymer structure will be flattened
by this effect.
Since the interactions among $x_{i\mu}$ through the branched polymer
structure is O($1/N$) suppressed (Note that there are 
only O($N$) number of bonds.), it is conceivable that the diagonal part of 
(\ref{decompose-diag-off}) is totally suppressed 
in the shrunken directions, and that in those directions,
one only sees the
fluctuation of the off-diagonal part $a_{\mu ij}$, which is 
mostly determined
by the Gaussian term (\ref{bosonic-OL}).
Since the coefficient $(\vec{x}_i-\vec{x}_j)^2$ 
in the Gaussian term (\ref{bosonic-OL})
is rotationally invariant, the fluctuation of $a_{\mu ij}$ is 
insensitive\footnote{The SO(6) breaking in the distribution of $\vec{x}_i$ 
may propagate to the fluctuation of $a_{\mu ij}$ 
through (\ref{detM}), but the effect is expected to be suppressed
by $1/N$.} to the SO(6) breaking in the distribution of $\vec{x}_i$.
This explains why we obtain similar extents in all shrunken directions
for each Ansatz.
Moreover since 
the magnitude of the coefficient $(\vec{x}_i-\vec{x}_j)^2$ is 
set by the typical length scale $\ell$ of the branched polymer,
it is also understandable that
the mean extent in the shrunken directions is universal
for all the SO($d$) symmetric vacuum with $d=3,4,5$.
The flattening of the branched polymer
may cause some $d$-dependence through (\ref{bosonic-OL})
in principle, but
the observed universality suggests that this
effect is small.

\begin{figure}[t]
\centerline{\includegraphics[width=.8\textwidth]{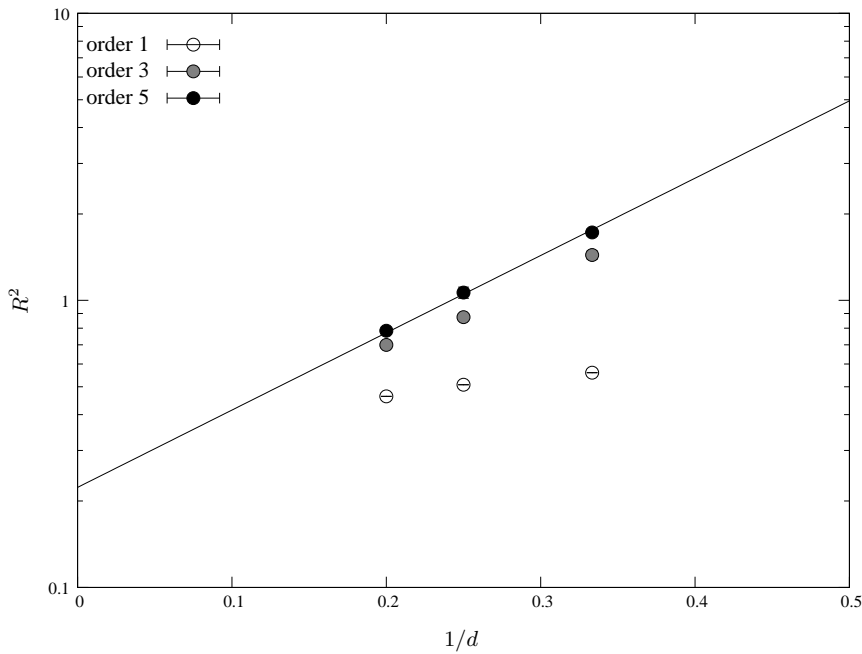}}
\caption{%
The extent of space-time $R^2$ in units of $g \sqrt{N}$
is plotted in the log scale against
$1/d$. The solid straight line is a linear fit to 
the results at order 5, 
from which we obtain $\tilde{r}^2/g \sqrt{N}=0.223$
and $\ell^2 /g \sqrt{N}= 0.627$ in Eq.~(\ref{predictingR}).
}
\label{fig:ext-R}
\end{figure}

Let us now turn our attention to the extent $R^2$ 
in the extended directions,
which we denote here as $(R_d)^2$ to
make its $d$-dependence manifest.
From entropic reasons, we may naively expect that
the branched polymer tends to occupy a
fixed volume in $D$ dimensions\footnote{Here we are discussing
the $D=6$ case, but we will see that (\ref{vol-indep}) actually
seems to hold also in the $D=10$ case.},
which implies that
\beq
(R_d)^{d} (\tilde{r})^{D-d} \approx \ell ^D \ ,
\label{vol-indep}
\eeq
where $\ell$ represents the extent of 
space-time in the model omitting the phase of the determinant.
The parameter $\tilde{r}$ 
represents the extent in the shrunken directions,
which we assume to be the same for all $d=3,4,5$
based on the universality discussed above, 
but leave its value to be unknown here.
This leads to
\beq
(R_d)^2  \approx  
\tilde{r}^2 \left(\frac{\ell^2}{\tilde{r}^2}\right) ^{D/d}  \ .
\label{predictingR}
\eeq
To test this behavior in the present $D=6$ case,
we plot in Fig.~\ref{fig:ext-R} 
the results for $(R_d)^2$ obtained by GEM (with $g^2 N=1$) 
in the log scale against $1/d$.
Indeed we find that the results tend to lie
on a straight line as expected from (\ref{predictingR})
as we increase the order.
{}By fitting the order 5 results to (\ref{predictingR}), 
we obtain\footnote{One can obtain these values graphically from
the straight line in Fig.~\ref{fig:ext-R} by noting that 
the intercept gives $\tilde{r}^2$
and the value of $R^2$ at $1/d=1/D$ gives
$\ell^2$.}
$\tilde{r}^2/(g\sqrt{N})=0.223$ and $\ell^2/(g\sqrt{N}) = 0.627$.
Note that the value of $\tilde{r}^2/(g\sqrt{N})$ 
obtained from this analysis turns out to be 
consistent with
the extent in the shrunken directions $r^2 \sim 0.2$ 
obtained by GEM directly with $g^2 N=1$, which is quite nontrivial.

We may consider $\ell^2/(g\sqrt{N})=0.627$ as a prediction from the
present argument. We can obtain it by Monte Carlo simulation
of the $D=6$ model omitting the phase of the fermion determinant.
The values of the $\lambda_i/(g\sqrt{N})$ ($1 \le i \le 6$)
are obtained up to $N = 32$,
and they all seem to converge to some value around 0.6
in the large-$N$ limit
assuming the finite-$N$ effects to be of O($1/N$)\cite{AAAHN}.
This also supports the validity of the constant volume property
(\ref{vol-indep}).

The extent in the shrunken directions 
being stable against increasing the order of the Gaussian expansion
can be understood from the fact that we are essentially seeing
the fluctuation of the off-diagonal elements, which are 
governed by an action like (\ref{bosonic-OL}).
On the other hand, the slow convergence of
the extent in the extended directions 
can be understood from the fact that we are essentially seeing
the fluctuation of the diagonal elements
governed by the branched-polymer dynamics.
The distribution of $\vec{x}_i$ tends to be 
uniform,
which is not well described by the Gaussian action. 
From this point of view, it is expected that
the convergence becomes slower for smaller $d$
since $(R_d)^2$ becomes larger according to (\ref{predictingR}).

\section{Results for the $\mbox{SO(2)}$ symmetric vacuum}
\label{sec:so2z4}

\begin{wrapfigure}{r}{6.6cm}
\centerline{\includegraphics[width=5.3cm]{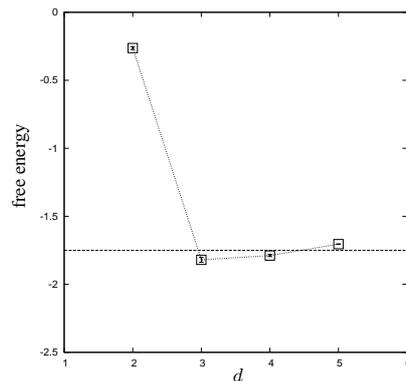}}
\caption{%
The free energy density for the SO($d$) symmetric vacuum 
obtained in this work is plotted against $d$. The horizontal dashed 
line represents the KNS value $f=-7/4$, and the dotted line connecting
the data points is drawn to guide the eye.}
\label{fig:free-d}
\end{wrapfigure}
In this Section we present some results for 
the SO(2) symmetric vacuum.
Since we have 13
parameters in the Gaussian action for the SO(2) symmetric vacuum,
we decided to 
impose $\mbox{SO(2)} \times \mathbb{Z}_4$ symmetry
as we discussed in Section \ref{sec:ansatz}.
Considering the results for the SO(3) symmetric vacuum
discussed in Section \ref{sec:result},
we expect that imposing the extra $\mathbb{Z}_4$ symmetry 
may not be that harmful.

In Table~\ref{tbl:resultSO2} we present the solutions of the 
self-consistency equations.
At order 5 we find two solutions (marked by asterisk
in the right-most column)
which have the extent in the shrunken directions
consistent with our result $r^2 \sim 0.2$ obtained 
for the SO($d$) symmetric vacua with $d=3,4,5$ universally.
The extent in the extended directions obtained for these
solutions is $(R_2)^2 = 2.3 \sim 2.6$, which 
is larger than those obtained for the SO(3) symmetric vacuum
at the same order.\footnote{If we use 
the formula (\ref{predictingR})
with $\tilde{r}^2=0.223$ and $\ell^2 = 0.627$,
we obtain $(R_2)^2 \sim 5$,
which is twice as large as the GEM result at order 5.
This might be due to the artifact of imposing 
the extra $\mathbb{Z}_4$ symmetry or due to the slow convergence
for large $R^2$ mentioned at the end of the previous Section.
}
Assuming that the universality extends to SO(2), 
we may consider these marked solutions at order 5
as the physical ones.
Then an estimate of the free energy density for 
the SO(2) symmetric vacuum is obtained as
$f=-0.25 \sim -0.27$ from Table~\ref{tbl:resultSO2},
which is considerably larger than the free energy density
for the SO($d$) symmetric vacuum with $d=3,4,5$.
This is consistent with
the argument that the 2d space-time is suppressed 
by the fermion determinant \cite{NV}.

\begin{table}[t]
\begin{center}
\begin{tabular}{c l  llll c}
\toprule
order & 
symmetry & 
\makebox[5em]{$f$} &
\makebox[3.5em]{$\langle \lambda_{1,2} \rangle$} &
\makebox[3.5em]{$\langle \lambda_{3,4,5} \rangle$} &
\makebox[3.5em]{$\langle \lambda_6 \rangle$} &
\\
\midrule
3 & $\SO(2)\!\times\!\mathbb{Z}_4$       & $ \phantom{-}2.29076$ & $ 0.13771$ 
& $0.82660$ 
& $ 1.60667$ & \\
\cline{3-7}
 &                                   & $ \phantom{-}1.66758$ & $ 1.32164$ 
& $0.18382$ 
& $0.15111$ & \\
\cline{3-7}
 &                                   & $ \phantom{-}0.10128$ & $ 1.06136$ 
& $0.20591$ 
& $0.42461$ & \\
\cline{3-7}
 &                                   & $ -2.35965$ & $ 1.81697$ 
& $0.26882$ 
& $0.48377$ & \\
\cline{3-7}
 &                                   & $ -2.42857$ & $1.72922$ 
& $0.21400$ 
& $0.44805$ & \\
\cline{3-7}
 &                                   & $-2.81818$ & $3.95383$ 
& $0.12571$ 
& $0.13949$ & \\
\cline{3-7}
 &                                   & $-3.27169$ & $5.85293$ 
& $0.08432$ 
& $0.10029$ & \\
\cline{3-7}
 &                                   & $-3.31790$ & $4.74086$ 
& $0.12446$ 
& $0.14714$ & \\
\hline
5 & $\SO(2)\!\times\!\mathbb{Z}_4$       & $ \phantom{-}2.13955$ & $0.14119$ 
& $0.87447$ 
& $1.72675$ & \\
\cline{3-7}
 &                                   & $ \phantom{-}1.51355$ & $1.51225$ 
& $0.18530$ 
& $0.15941$ & \\
\cline{3-7}
 &                                   & $ \phantom{-}0.16798$ & $2.83645$ 
& $0.12589$ 
& $0.13595$ & \\
\cline{3-7}
 &                                   & $ -0.13712$ & $3.18237$ 
& $0.12008$ 
& $0.11814$ & \\
\cline{3-7}
 &                                   & $ -0.25243$  & $2.29020$ 
& $0.16355 $ 
& $0.17833 $ & $\ast$ \\
\cline{3-7}
 &                                   & $-0.27313$  & $2.57212$ 
& $0.14915$ 
& $0.15902$ & $\ast$ \\
\cline{3-7}
 &                                   & $-1.18723$  & $6.87681$ 
& $0.04775 $ 
& $0.03738 $ & \\
\cline{3-7}
 &                                   & $-1.22564 $  & $7.25959 $ 
& $0.04115 $ 
& $0.01773 $ & \\
\cline{3-7}
 &                                   & $-2.45562$  & $11.8953$ 
& $0.03757$ 
& $0.04213$ & \\
\cline{3-7}
 &                                   & $-2.63861$  & $12.1413$ 
& $0.03970 $ 
& $0.04170 $ & \\
\cline{3-7}
 &                                   & $-2.92543$  & $13.7449$ 
& $0.03624$ 
& $0.03723 $ & \\
\cline{3-7}
 &                                   & $-3.55149$  & $0.42405$ 
& $0.46966$ 
& $0.45653$ & \\
\cline{3-7}
 &                                   & $-3.57951$  & $8.75022$ 
& $0.05582$ 
& $0.06260$ & \\
\cline{3-7}
 &                                   & $-3.6158$  & $0.29921$ 
& $0.73480$ 
& $0.36202$ & \\
\cline{3-7}
 &                                   & $-3.90928$  & $9.94824$ 
& $0.04923$ 
& $0.05356$ & \\
\cline{3-7}
 &                                   & $-4.11550$  & $5.17586$ 
& $0.10051$ 
& $0.10911$ & \\
\cline{3-7}
 &                                   & $-4.20561$  & $6.79049$ 
& $0.07689 $ 
& $0.08446 $ & \\
\cline{3-7}
 &                                   & $-12.1677 $  & $0.19044$ 
& $1.53911$ 
& $1.77276$ & \\
\bottomrule
\end{tabular}
\end{center}
\caption{%
Numerical values of the free energy density
and the eigenvalues 
of the moment-of-inertia tensor evaluated at the solutions 
obtained for the $\mbox{SO(2)} \times \mathbb{Z}_4$ Ansatz.
Asterisks in the right-most
column indicate the solutions we consider to be ``physical''.
}
\label{tbl:resultSO2}
\end{table}

In Fig.\ \ref{fig:free-d} we plot 
the free energy density\footnote{The behavior 
of the free energy density for $d=2,3,4,5$
is analogous to the results for the $D=10$ case
obtained at the 3rd order \cite{Nishimura:2001sx}
for $d=2,4,6,7$, where $d=4$ is found to give the minimum.} 
for the SO($d$) symmetric vacuum against $d$.
(When there are more than one ``physical solutions''
we take the average and put an error bar representing
the mean square error as in Fig.\ \ref{fig:rr}.)
Thus we conclude that
the SO(3) symmetric vacuum gives the smallest free energy,
and hence it is chosen as the true vacuum.

\section{Reconsideration of the $D=10$ case}
\label{sec:10dcase}

\begin{figure}[H]
\begin{center}
\begin{tabular}{cc}
\epsfig{file=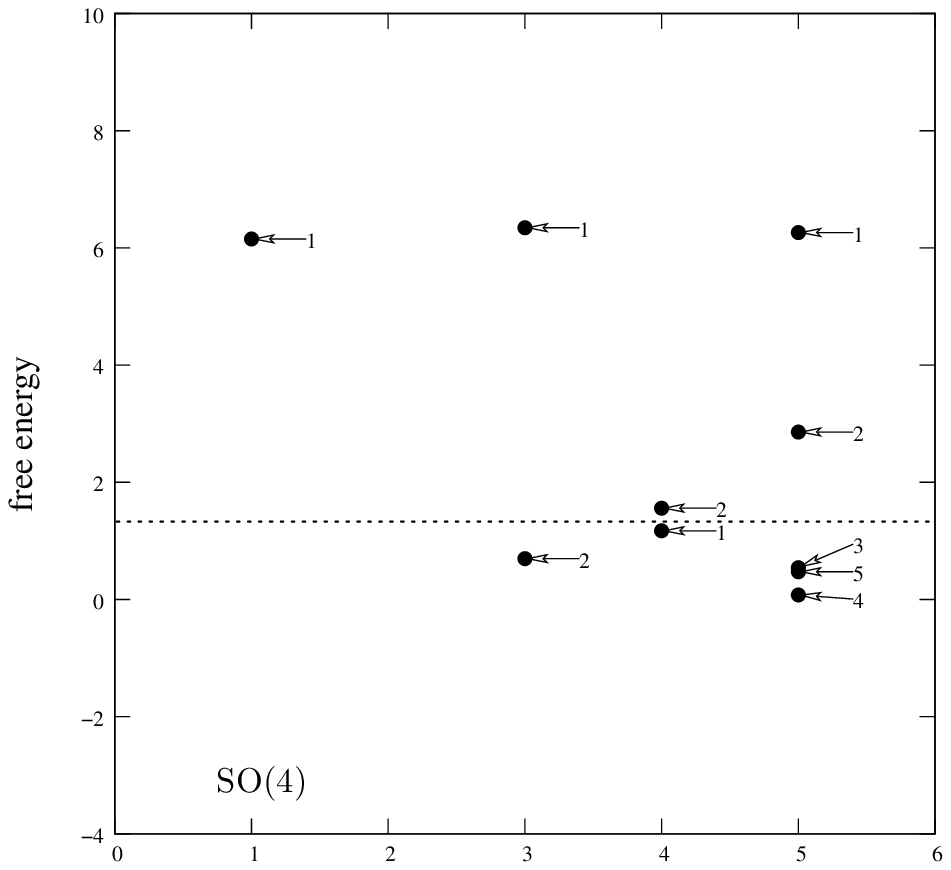,width=.47\textwidth} &
\epsfig{file=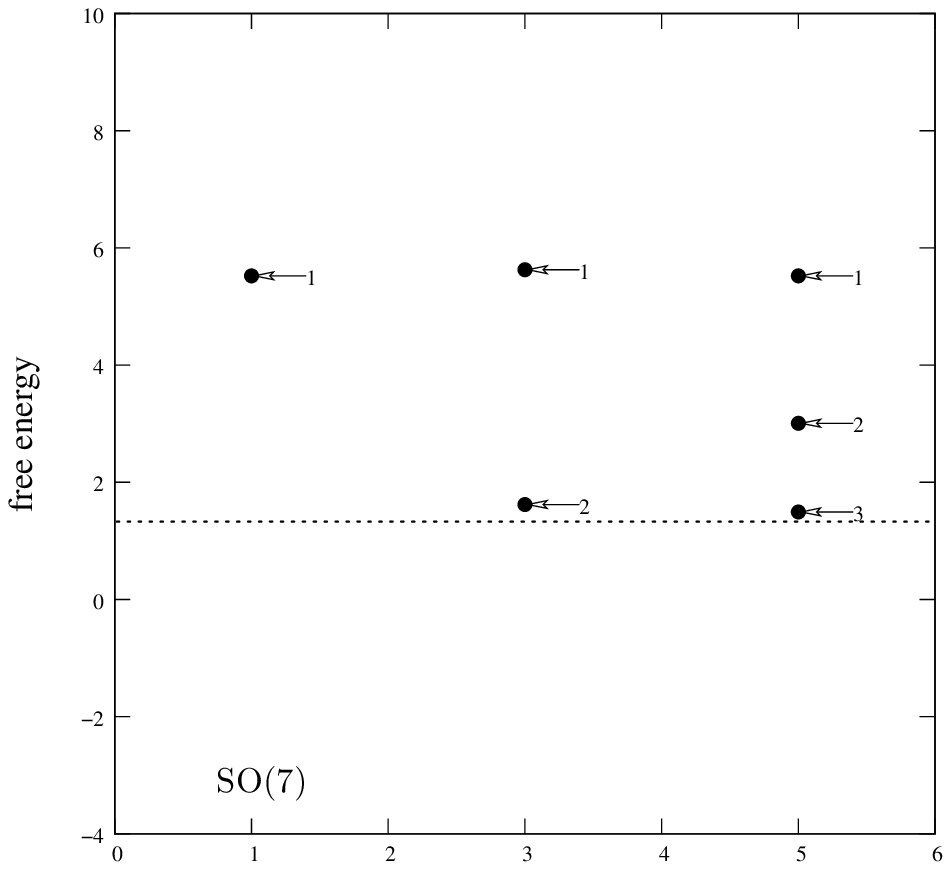,width=.47\textwidth} \\
\epsfig{file=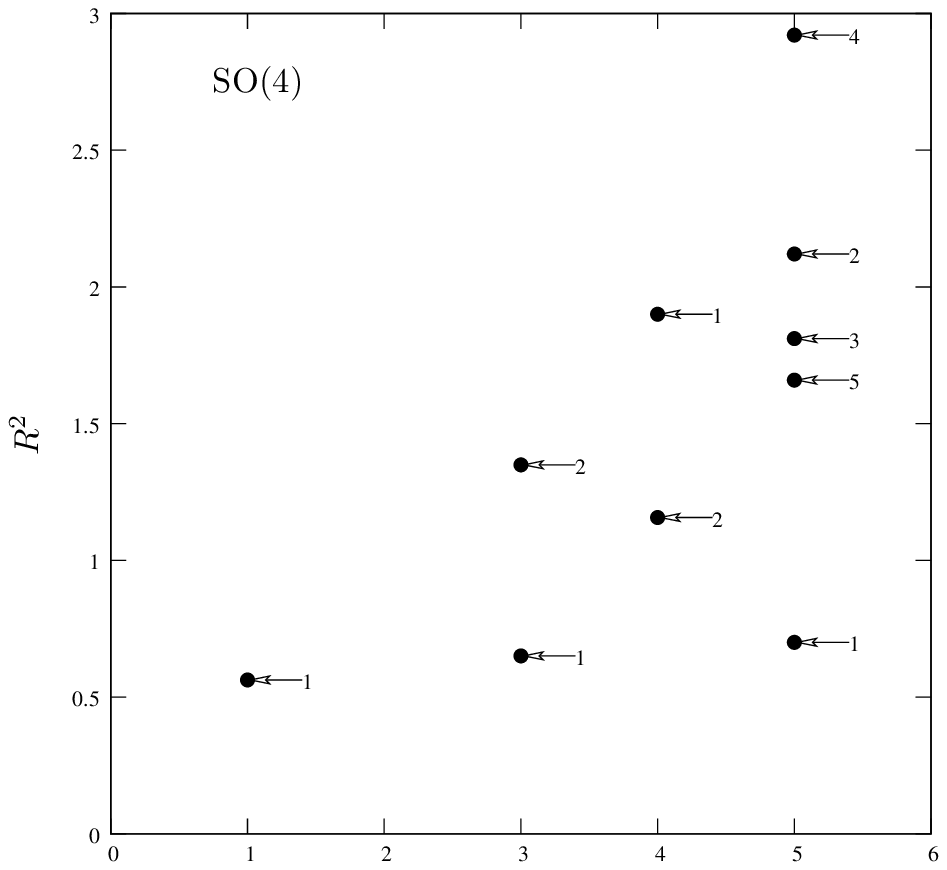,width=.47\textwidth} &
\epsfig{file=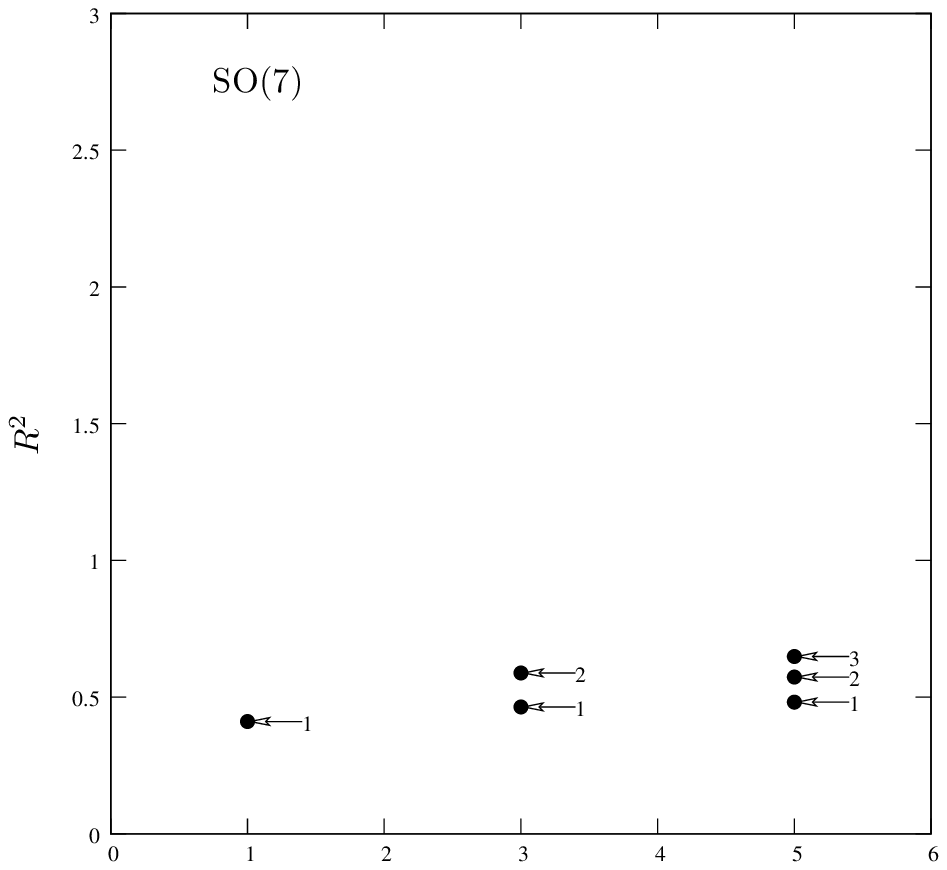,width=.47\textwidth} \\
\epsfig{file=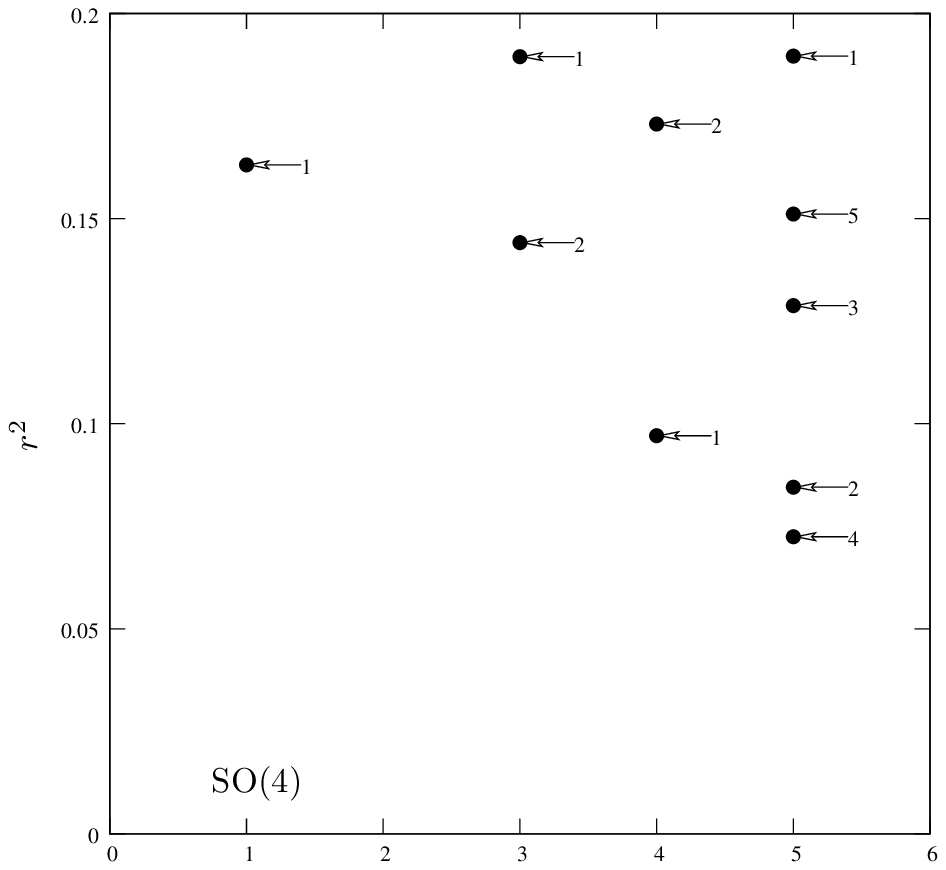,width=.47\textwidth} &
\epsfig{file=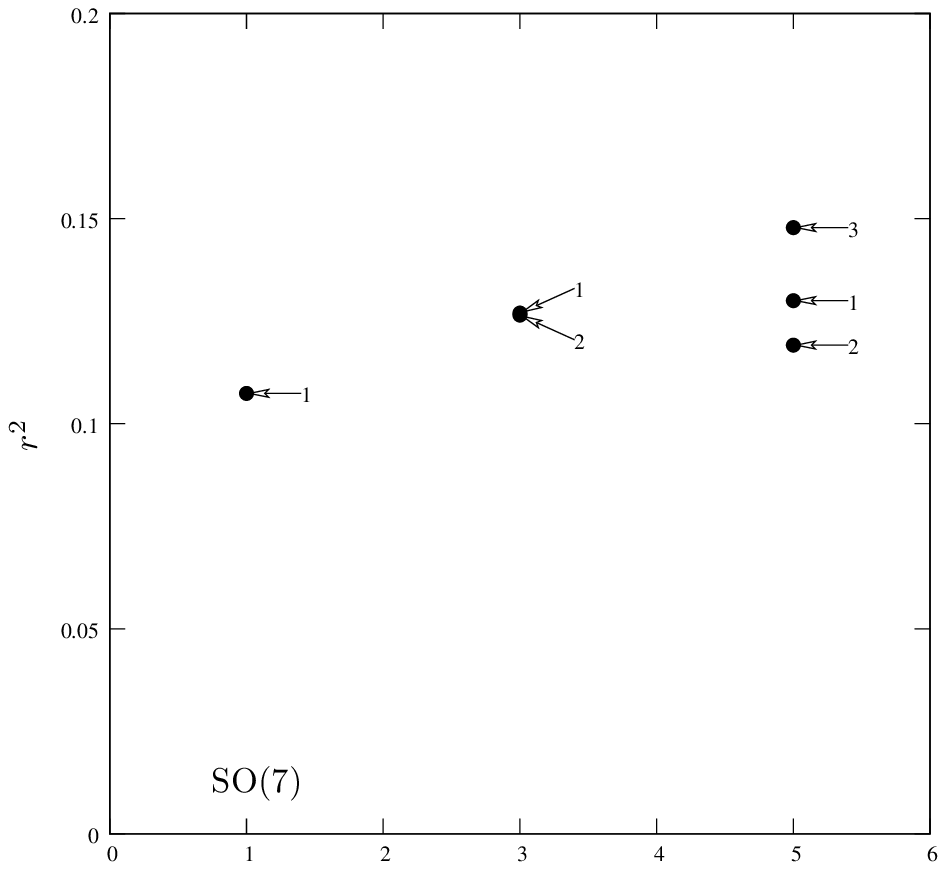,width=.47\textwidth} \\
\end{tabular}
\end{center}
\caption{%
The results for the $D=10$ model obtained by GEM
up to the 5th order 
taken from Ref.~\citen{Kawai:2002ub}. 
The free energy density (top), 
the extent of space-time
$R^2$ in the extended directions (middle) and 
$r^2$ in the shrunken directions (bottom)
are plotted against the order of GEM
for the $\SO(4)$ Ansatz (left column) 
and the $\SO(7)$ Ansatz (right column) with
additional symmetries described in the text.
The horizontal dotted lines in the top panels
represent the KNS value
($\ln 8 - \frac{3}{4}=1.32944$)
for the $D=10$ model obtained in Ref.~\citen{Nishimura:2001sx}.
The arrows with a number are added to 
identify solutions obtained at each order.
}
\label{fig:ikkt}
\end{figure}

In this Section we 
use the new insights obtained from the study of the $D=6$ case
to reconsider the previous results
obtained by GEM in the $D=10$ case.
Fig.~\ref{fig:ikkt} shows the results
for the $D=10$ model 
up to the 5th order\footnote{For the $D=10$ model,
there are results also at 7th \cite{Kawai:2002ub}
and 8th \cite{Aoyama:2006rk} orders. 
(Original calculations were done at 
the 3rd order \cite{Nishimura:2001sx}.)
For the SO(7) symmetric vacuum,
the free energy density and the extent of space-time
are reasonably stable against increasing the order.
On the other hand, for the SO(4) symmetric vacuum,
higher order calculations give no solutions in the region where
we identified as the location of a 
plateau from the results at the 5th order.
For instance, there are no solutions which give the extent
in the shrunken directions around $0.13\sim 0.15$.
Considering the nice theoretical consistencies we find below,
we suspect that higher order calculations for the SO(4) symmetric vacuum
have some problems possibly 
due to the imposed extra symmetries or due to the bad convergence
for large $R^2$ mentioned at the end of Section \ref{sec:theo-interpret}.}
taken from Ref.~\citen{Kawai:2002ub}.
In order to reduce the number of arbitrary parameters,
two types of symmetry were imposed.
One is the $\SO(7)\!\times\!\SO(3)$ symmetry,
and the other is 
the $\SO(4)\!\times\!\SO(3)\!\times\!\SO(3)\!\times\!\mathbb{Z}_2$ 
symmetry, where $\mathbb{Z}_2$ corresponds to interchanging 
the two $\SO(3)$.

For the SO(4) 
Ansatz, 
we find that solutions No.3 and No.5
at order 5 give similar values for all the quantities, 
which we consider as an indication of the plateau formation.
For the SO(7) 
Ansatz,
we take the solution No.3 to be the physical one
considering the similarity of the pattern to
what we have observed for the SO(5) symmetric vacuum in the $D = 6$ model;
See Fig.~\ref{fig:f}.
Note also that the values of free energy density
obtained for the ``physical solutions'' we picked up above
are quite close to the KNS value represented by the
horizontal dotted line.

Then we find that the situation
is actually quite similar to the $D=6$ case.
First the free energy density is smaller for the SO(4) 
symmetric vacuum than for the SO(7) 
symmetric vacuum.
The extent of space-time 
in the shrunken directions is around $r^2 = 0.13\sim 0.15$
for both 
vacua.
On the other hand, the extent of space-time in the extended directions
is estimated as $(R_4)^2 = 1.6 \sim 1.8$ for the SO(4) 
symmetric vacuum
and $(R_7)^2 = 0.52 \sim 0.65$ 
for the SO(7) symmetric vacuum.\footnote{Here
the upper bound 0.65 is taken from the solution No.3 at the 5th order,
and the lower bound 0.52 is taken 
from the solution at the 7th order \cite{Kawai:2002ub},
which gives results closest to the solution No.3 at the 5th order.
}

\begin{figure}[t]
\centerline{\includegraphics[width=.8\textwidth]{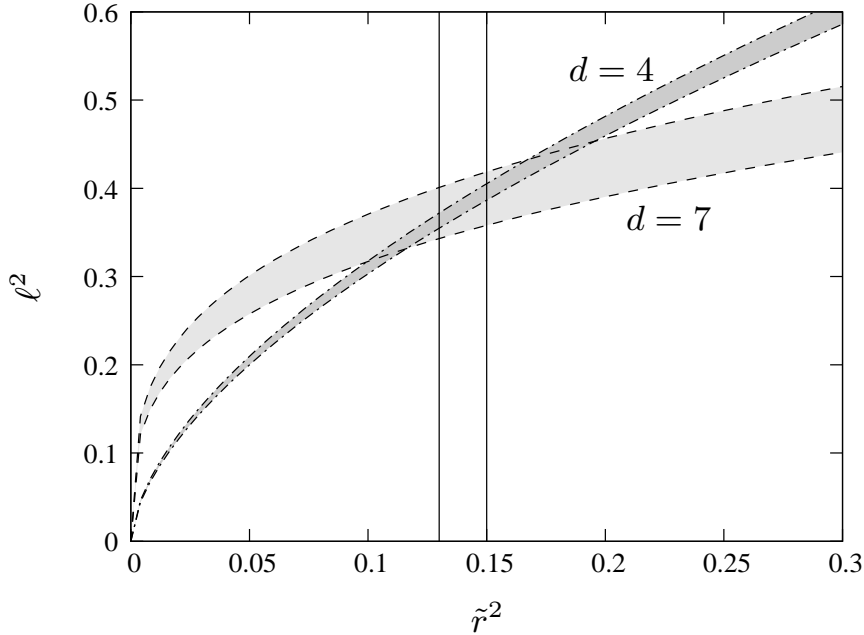}}
\caption{%
Constraints on $\tilde{r}^2$ and $\ell^2$ from 
the formula (\ref{vol-indep})
with the input of the extent of space-time 
in the extended directions obtained by GEM
in the $D=10$ case for the SO($4$) symmetric vacuum 
(the region between the two dash-dotted lines)
and the SO($7$) symmetric vacuum 
(the region between the two dashed lines).
The vertical solid lines represent the range of $r^2$
(the extent of space-time in the shrunken directions) 
obtained by GEM directly.
There is a region in which all the three constraints are 
satisfied.
}
\label{fig:three-constraints}
\end{figure}

Assuming the formula (\ref{vol-indep}) to hold also in the $D=10$ model
and using the values of $(R_d)^2$ extracted above,
we can put constraints on $\ell^2$ and $\tilde{r}^2$
as shown in Fig.~\ref{fig:three-constraints}.
We also indicate the region of $\tilde{r}^2$
suggested from the extent of space-time 
in the shrunken directions $r^2=0.13\sim 0.15$
obtained directly by GEM.
Quite nontrivially, we find a region
in which all the three constraints are satisfied.
The allowed region for $\ell^2$ turns out to be
$\ell^2=0.35 \sim 0.4$.
As in the $D=6$ case, we expect that this value can be
reproduced by calculating the extent of space-time
in the model omitting the phase of the fermion determinant.
Preliminary results of Monte Carlo simulation for $D=10$ suggest
that this is indeed the case \cite{AAAHN}.
We therefore consider that the values of observables extracted above
from the physical solutions 
are sensible.
Note, in particular, that we have $R_4/r=3.3 \sim 3.7$,
which is finite as opposed to previous speculation that the
ratio might be infinite from the results 
at higher orders \cite{Aoyama:2006rk}.
Plugging $r^2=0.13\sim 0.15$ and $\ell^2=0.35 \sim 0.40$
into Eq.~(\ref{predictingR}), 
we can also predict the values of $(R_d)^2$ for $d$ other than $d=4,7 $.

Thus we think that we have arrived at comprehensive
understanding on the dynamics of 
``space-time''
in dimensionally reduced super Yang-Mills models for
both $D=6$ and $D=10$.

\section{Summary and discussions}
\label{sec:discussion}

We have discussed the SSB of SO($D$) in
the large-$N$ reduced models obtained by 
the zero-volume limit of 
pure SU($N$) super Yang-Mills theories in $D=6,10$ dimensions.
Both models have a complex fermion determinant,
which is expected to play a crucial role in 
the conjectured SSB \cite{NV,Nishimura:2001sq,Nishimura:2004ts,sign}.

First we studied the $D=6$ model
by GEM
up to the 5th order.
Unlike the previous studies of the $D=10$ model, we were able to examine
the SO($d$) symmetric vacua with $d=3,4,5$ without imposing 
{\em ad hoc} symmetries in the extra dimensions in most cases. 
From a set of solutions to the self-consistency equations
giving results close to each other, 
clear indication of the plateau formation was observed.
This enabled us to obtain reliable results for the
free energy and the extent of space-time in each direction
as in an analogous study of the toy model \cite{Nishimura:2004ts}.

We found that the free energy decreases 
as we go from $d=5$ to $d=3$.
In fact it was argued in Ref.~\citen{Aoki:1998vn} that
a branched polymer is difficult to collapse
to a hypersurface with less than four dimensions, 
since the Hausdorff dimension of such a system is four.
The reason why our conclusion can still be true is that
the system has a finite extent in the shrunken directions
for fixed $g^2 N$.
Namely the branched polymer is not really collapsed to a hypersurface.

In fact we found that the shrunken directions have 
approximately the same extents for $d=3,4,5$.
We have given a theoretical explanation of this behavior
based on the low-energy effective theory.
On the other hand, the extended directions have 
a larger extent for smaller $d$,
and this $d$-dependence can be nicely
explained by the law of constant volume (\ref{vol-indep}).
From this observation, we obtained a prediction for the extent of
space-time in the phase-quenched model 
(\emph{i.e.}, the model obtained by 
omitting the phase of the fermion determinant),
which seems to be consistent with some preliminary
results obtained by Monte Carlo simulation.

The results for the SO(2) symmetric vacua are not clear
possibly due to the extra symmetries we had to impose
in the extra dimensions. However, if we assume that
the extent of space-time in the shrunken directions
is close to what we obtained for $d=3,4,5$,
we obtain the value of free energy much larger than 
the values obtained for $d=3,4,5$.
This is consistent with
the suppression of two-dimensional space-time
due to the fermion determinant \cite{NV}.
Thus we conclude that the SO(6) symmetry of the $D=6$ model
is spontaneously broken down to SO(3).

The new insights obtained from the $D=6$ model
enabled us to reinterpret 
the previous results for the $D=10$ model.
In particular, we considered the results 
for the SO(4) and SO(7) symmetric vacua
obtained at the 5th order.
We found that the extent in the shrunken directions is similar
for the two vacua. 
Assuming the constant volume property (\ref{vol-indep}) 
and using the extent 
in the extended directions as an input, we obtained the possible 
region for $\tilde{r}^2$ and $\ell^2$. 
It turned out that the value of $\tilde{r}^2$ suggested by this 
analysis is consistent with the extent in the shrunken 
directions obtained by GEM directly.
The value of $\ell^2$, on the other hand, provides a prediction for 
the extent of the space-time 
in the phase-quenched model, which is also consistent with
some preliminary results of Monte Carlo simulation.

The free energy in the $D=10$ model was calculated 
for $d=2$, $4$, $6$, $7$ at the 3rd order, and 
$d=4$ was found to give the minimum.\cite{Nishimura:2001sx} 
The particular pattern of the $d$-dependence obtained at this order
might already capture the correct qualitative behavior
considering the results obtained at the 5th order 
for $d=4$ and $d=7$.
This $d$-dependence is similar to what we obtained
in the $D=6$ model up to the 5th order, 
where $d=3$ gives the minimum. 
In order to determine the true vacuum of the $D=10$ model,
we therefore consider it important
to compare the free energy 
for $d=3$, $4$, $5$, 
including the $d=3$ and $5$ cases which have not been studied
so far.

As is discussed in Refs.~\citen{NV,Nishimura:2001sq,Nishimura:2004ts},
the SSB of SO($D$) is caused by the phase of the fermion determinant.
Rotationally symmetric configurations are strongly suppressed
due to cancellations caused by the violent fluctuation of the phase,
whereas the fluctuation becomes milder for collapsed configurations.
Monte Carlo studies of these models are difficult
precisely because of the cancellations.
However, a new method to sample efficiently the dominant
configurations including the effects of the phase
is proposed in Ref.~\citen{sign}.
It is interesting to study 
the $D=6$ and $D=10$ models by the Monte Carlo 
method
to see whether the results and predictions
obtained in this paper can be reproduced.
We expect that the method will eventually enable us to
reach a definite conclusion on which value of $d$
is chosen dynamically in the $D=10$ model.

To conclude, we consider it interesting that
the SSB of SO(6) down to SO(3) has been demonstrated
clearly in the $D=6$ model.
Our results make it very plausible that
an analogous SSB occurs 
in the $D=10$ model,
and that the SO(10) symmetry is broken down to 
either SO(3), SO(4) or SO(5).
This strongly supports
the speculation that 
the IIB matrix model may provide
the dynamical 
origin of the space-time dimensionality.
The space-time picture that emerged
from the present work is also interesting.
The extra dimensions are totally dominated by fluctuations of 
off-diagonal elements,
and presumably do not
allow ordinary geometric descriptions.
This may provide certain basis for
phenomenological models \cite{Chatzistavrakidis:2010xi}
with non-commutative extra dimensions.
The space-time in the extended directions is uniform
and it is described mostly by the 
commutative degrees of freedom.

On the other hand, the constant volume property (\ref{vol-indep})
suggested by our results implies that
the ratio $R/r$ of the extent of space-time 
in the extended directions and that in the shrunken directions
is finite as opposed to previous speculations that
it might be infinite \cite{Aoyama:2006rk}.
We feel that this gives us an important clue on how we should actually
interpret the IIB matrix model as 
a non-perturbative definition of 
type IIB superstring theory.\footnote{Let us note here that 
the identification of the string scale in the IIB matrix model 
is also an important open question.
For instance, the correlation functions of Wilson loops
in the $D=4$ version of the IIB matrix model
were measured by Monte Carlo simulation \cite{Ambjorn:2000bf},
and the large-$N$ scaling behavior was observed with
$g^2 N$ fixed.
This must be true also 
in the $D=6$ and $D=10$ models since the extent of space-time
in the extended directions
has qualitatively the same large-$N$ behavior as the $D=4$ model
as we have revealed in this paper.
Then one has to identify $l_{s}=\sqrt{g}N^{1/4}$ as the string
scale. The same conclusion was obtained
in an attempt to derive the string field Hamiltonian
from the IIB matrix model \cite{Fukuma:1997en}.
However, there are also different proposals
based on some other arguments \cite{Aoki:1998vn,Kawai:2007tn}.}
A possible interpretation would be that the model 
actually describes the state of the
early universe in the spirit of Hartle-Hawking \cite{Hartle:1983ai}
as suggested by the fact that the model
uses the Euclidean signature for the space-time \cite{Kaneko:2005kp}.
From this point of view, 
it might be interesting to consider a matrix model
which incorporates the Lorentzian signature for the space-time,
or a matrix model which includes the time
from the outset
(instead of generating it dynamically)
as in the BFSS matrix model \cite{Banks:1996vh}.
In any case we hope that our findings in this paper motivate
further investigations of the matrix model approach to 
non-perturbative aspects of superstring theories
and the origin of our space-time.

\section*{Acknowledgements}
We would like to thank Konstantinos Anagnostopoulos,
Takehiro Azuma, Hajime Aoki,
Masanori Hanada, Satoshi Iso, Hikaru Kawai,
Yoshihisa Kitazawa, Fumihiko Sugino,
Shun'ya Mizoguchi and Asato Tsuchiya for discussions.
The work of J.N.\ is supported in part by Grant-in-Aid 
for Scientific Research 
(No.\ 19340066 and 20540286)
from Japan Society for the Promotion of Science.

\appendix

\section{Symplectic Majorana-Weyl spinors in 6d}
\label{sec:symplectic}

In $D=6$ one cannot define Majorana spinors, but 
one can define symplectic Majorana spinors, 
and impose the Weyl condition simultaneously. 

Let $\psi$ be a Weyl spinor with positive chirality.
Its charge conjugation $\psi_{\rm c}$ can be defined by
\begin{equation}
  \psi_{\rm c} = C^{-1} \bar{\psi}^T \ , \label{eq:cc}
\end{equation}
where $C$ is the charge conjugation matrix satisfying
$C\Gamma_\mu C^{-1} = (\Gamma_\mu)^T$ and $C^T = -C$.
Then we define symplectic Majorana-Weyl spinors as
\begin{eqnarray}
  \psi_1 &=& \frac{1}{2} (\psi - \psi_{\rm c}) \ , \label{eq:smw1} \\
  \psi_2 &=& \frac{1}{2} (\psi + \psi_{\rm c}) \ . \label{eq:smw2}
\end{eqnarray}
It is conventional to define
\begin{equation}
  \psibar^{\, i} = \psi_i^\dagger \Gamma^0 \quad (i=1,2) \ .
\end{equation}
In this notation, we have
\begin{equation}
  ( \psibar^i )^T = C \psi^i  \ ,
\end{equation}
where
\begin{equation}
  \psi^i = \epsilon^{ij} \psi_j \ , 
\quad \psi_j = \psi^i \epsilon_{ij} \ , \quad
  \epsilon^{12} = \epsilon_{12} = +1 \ .
\end{equation}
The symplectic indices are raised and lowered 
by contracting with $\epsilon^{ij}$
and $\epsilon_{ij}$, respectively, according to the NW-SE rule.

Let us decompose fermionic variables $\Psi_\alpha$ in
(\ref{eq:sf}) into 
\begin{equation}
  \Psi_\alpha = \Psi_{1,\alpha} + \Psi_{2,\alpha} \ , 
\label{eq:Phi_smw}
\end{equation}
where $\Psi_{1,\alpha}$, $\Psi_{2,\alpha}$ are the components of
symplectic Majorana-Weyl spinor.
Then, the fermionic part of the original action (\ref{eq:sf}) 
and the Gaussian action (\ref{eq:s0f}) can be rewritten, respectively, as
\begin{eqnarray}
  S_{\rm f} &=& N \Tr (\Psi_{i,\alpha} 
\epsilon^{ij} (C \Gamma^\mu)_{\alpha\beta}
    [ X_\mu, \Psi_{j,\beta} ]) \ , 
\label{eq:sf_smw} \\
  S_{\rm 0f} &=&
    N \sum_{\alpha,\beta = 1}^{4} \mathcal{A}^{ij}_{\alpha\beta}
  \Tr ( \Psi_{i,\alpha} \Psi_{j,\beta} )  \ , 
\label{eq:s0f_smw} \\
 \mathcal{A}^{ij}_{\alpha\beta} &=& (\sigma_3)^{ij}  
    \sum_{\mu = 1}^{6} 
    m_\mu (C \Gamma_\mu)_{\alpha\beta}
    + \frac{i}{2\cdot 3!} \epsilon^{ij} \!\!\!\!
  \sum_{\mu,\nu,\rho = 1}^{6} m_{\mu\nu\rho}
    (C \Gamma_\mu \Gamma_\nu^\dagger \Gamma_\rho)_{\alpha\beta} \ ,
\label{eq:decomp_m_smw}
\end{eqnarray}
where $m_\mu$ and $m_{\mu\nu\rho}$ are 
the parameters introduced in (\ref{eq:decomp_m}).

By using the symplectic Majorana-Weyl spinors
instead of Weyl spinors, 
one can get rid of the orientation of
the fermion propagators,
and therefore the number of Feynman diagrams is reduced
considerably.
In fact the list of Feynman diagrams to be considered
is exactly the same as in the IIB matrix model.
The summation over the symplectic indices for each diagram
can be done very easily.

\section{Free energy from the Krauth-Nicolai-Staudacher conjecture} 
\label{sec:KNSresult}

In this Appendix we describe the analytic formula for the partition function
conjectured by Krauth, Nicolai and Staudacher (KNS) \cite{Krauth:1998xh}
combining their Monte Carlo results at small $N$ \cite{Krauth:1998xh}
and earlier analytic works \cite{Green:1997tn,Moore:1998et}.
For the present model, the formula reads
\begin{eqnarray}
  Z_{\rm KNS} &=& \int \dd A \, \dd \Psi \, \dd \Psibar
\, e^{- S_{\rm KNS}} \  \nonumber \\
  &=&  \dfrac{2^{\frac{N(N+1)}{2}} \pi^{\frac{N-1}{2}}}
	{2\sqrt{N}\prod_{k=1}^{N-1} k!}
  \times
  \frac{1}{N^2} \ ,
  \label{eq:analytic-kns} \\
  S_{\rm KNS} &=& \frac{2}{N}(S_{\rm b} + S_{\rm f}) \ ,
\end{eqnarray}
where $S_{\rm b}$ and 
$S_{\rm f}$ are defined by (\ref{eq:sb}) and (\ref{eq:sf}), respectively.
In the above formula, the definition of the action $S_{\rm KNS}$
differs from our definition by the factor of $2/N$. 
In order to absorb this factor, we introduce the rescaled variables 
$A'_\mu = (2/N)^{1/4} A_\mu$,
$\Psi'_\alpha = (2/N)^{3/8} \Psi_\alpha$
and $\bar{\Psi}'_\alpha = (2/N)^{3/8} \bar{\Psi}_\alpha$,
whose integration measure 
is given by
\begin{equation}
\dd A' \, \dd \Psi' \, \dd \Psibar' =
 \left( \frac{N}{2} \right)^{\frac{3}{2}(N^2-1)} 
\dd A \, \dd \Psi \, \dd \Psibar \ .
\end{equation}
As a result, the partition function
(\ref{eq:6dpf}) can be obtained as
\begin{eqnarray}
  Z &=& \left(\frac{N}{2}\right)^{\frac{3}{2}(N^2 - 1)} 
Z_{\rm KNS} \nonumber \\
  &=& 2^{-N^2 -\frac{N}{2} 
-\frac{5}{2}} \pi^{\frac{N-1}{2}} N^{\frac{3}{2}N^2 - 4} 
\prod_{k=1}^{N-1} k!   \  .
\end{eqnarray}
{}From this, we obtain the large-$N$ asymptotics
as
\begin{equation}
  \frac{F}{N^2 - 1} = 
  - \ln N + \ln 2 - \frac{3}{4}
  + O\left( \frac{\ln N}{N^2} \right) \ .
  \label{eq:analytic}
\end{equation}
GEM
reproduces the first term correctly 
for any Ansatz.
Substituting this into the definition (\ref{eq:fedensity}) of 
the ``free energy density'',
we obtain $f=-\frac{7}{4}=-1.75$ as a prediction from the KNS conjecture.


\end{document}